\documentclass[a4paper,twocolumn,11pt,unpublished=2026-08-20]{quantumarticle}
\pdfoutput=1
\usepackage[utf8]{inputenc}
\usepackage[english]{babel}
\usepackage[T1]{fontenc}
\usepackage{amsmath}
\usepackage{braket}
\usepackage{qcircuit}
\usepackage{tikz}
\usepackage{lipsum}
\usepackage{ulem, mathtools}
\usepackage{xurl}
\usepackage{hyperref}
\usepackage{subcaption}

\usepackage{etoolbox}
\usepackage{csquotes}

\usepackage[backend = biber, style = ieee]{biblatex}
\usepackage[section]{placeins}
\usepackage{float}
\usepackage{xcolor}

\begin{document}

\title{Benchmarking Quantum Simulations of the Lipkin-Meshkov-Glick Model Using Large Tensor Networks}
\author{Maggie Bao}
\affiliation{The Washington Institute for STEM Entrepreneurship and Research (WISER)}
\affiliation{Massachusetts Institute of Technology, 77 Massachusetts Avenue, Cambridge, MA 02139,
USA}
\orcid{0000-0003-1533-8015}

\author{Rushil Dandamudi}
\affiliation{The Washington Institute for STEM Entrepreneurship and Research (WISER)}
\affiliation{Joint Center for Quantum Information and Computer Science, University of Maryland, College Park, MD 20742, USA}
\orcid{0000-0003-1985-4623}

\author{Jerimiah Wright}
\affiliation{The Washington Institute for STEM Entrepreneurship and Research (WISER)}
\orcid{0000-0003-2886-9210}

\author{Joan \'Etude Arrow}
\affiliation{Naval Nuclear Laboratory, 2401 River Rd, Niskayuna, NY, 12309, USA}
\orcid{0009-0004-3176-5409}


\author{Henry Zou}
\affiliation{IBM Quantum, IBM Research, Cambridge, MA 02142, USA}
\orcid{0000-0002-0335-9508}

\author{Vardaan Sahgal}
\affiliation{The Washington Institute for STEM Entrepreneurship and Research (WISER)}
\orcid{0000-0002-2293-7430} 

\author{Brian J. McDermott}
\affiliation{Naval Nuclear Laboratory, 2401 River Rd, Niskayuna, NY, 12309, USA}
\affiliation{Rensselaer Polytechnic Institute, 110 8th Street, Troy, NY 12180}
\orcid{0000-0002-0335-9508}
\maketitle

\begin{abstract}

As quantum computing matures, it is critical to benchmark its real-world problem solving performance against competitive classical methods, such as tensor networks. In this work, we leverage the Density Matrix Renormalization Group (DMRG) algorithm to compute ground state energies of the Lipkin–Meshkov–Glick (LMG) model as a comparative benchmark against popular noisy intermediate-scale (NISQ) algorithms like the Variational Quantum Eigensolver (VQE) and Sample-Based Quantum Diagonalization (SQD) method. By running DMRG on the NERSC Perlmutter supercomputer, we provide one of the largest LMG ground state energy datasets in literature, containing accurate ground-state energies for systems up to 1400 particles. We compare these results with VQE and SQD implementations on an IBM Eagle quantum computer for comparison. VQE achieved results within 1-percent error for 6 particles, while exceeding that threshold for all other values while SQD extended that range to 17 particles, suggesting that in a Noisy Intermediate-Scale Quantum era, subspace-based approaches may strike the best balance between accuracy, circuit depth, and noise resilience.  

\end{abstract}

\section{Introduction}

Quantum computing promises a novel approach to the simulation of quantum systems beyond the capabilities of traditional computational methods \cite{feynman_simulating_1982}. Despite this promise, Noisy Intermediate-Scale Quantum (NISQ) devices do not yet demonstrate clear and unambiguous advantages over classical computing \cite{mind_the_gaps_2025}. However, as improvements to NISQ hardware and algorithms yield utility-scale performance, improved methodologies that benchmark quantum computation on key applications of interest must be developed. Historically, quantum benchmarking has focused on low-to-medium level hardware characterizations such as coherence time, gate fidelity, or quantum volume \cite{cross_validating_2019}. While these benchmarks are vital for cross-comparing hardware, they do not fully capture how well quantum computers may perform on various applications of real-world interest. More recently, application-centered benchmarks for quantum computers have begun to emerge \cite{lubinski_optimization_2024, butko_hamperf_2024}. Rather than centering the hardware platforms themselves, application-based quantum benchmarks allow multiple hardware platforms to be assessed in terms of their ability to solve key problems of interest to end-users of the technology. Each application of quantum computers may come with multiple problem Hamiltonians that can be used to encode such an application. The Hamiltonian Library (HamLib) \cite{sawaya_hamlib_2024} encodes various applications of interest into the Hamiltonian used by the quantum computer to solve the problem. The models contained in HamLib cover a broad spectrum of quantum applications, including chemistry, optimization, and condensed matter. Building on such standardized Hamiltonian problem sets, the Hamiltonian Performance (HamPerf) benchmarking framework established in \cite{butko_hamperf_2024} defines a method for cross comparison of quantum algorithms and hardware against more conventional classical methods. This comparison is vital for clarifying the value proposition of quantum methods relative to more standard methodologies in industry. Developing an expanded set of application benchmarks, therefore, remains a key area of interest for emerging algorithms. In particular, we aim to contribute a benchmark of interest to the nuclear engineering community to the framework provided by HamPerf and HamLib. This framework identifies a benchmark based on the Hamiltonian $\hat{H}$ or by a dataset of ground state energies of $\hat{H}$ that can be compared against the output of a particular algorithm/hardware pair under evaluation.

In this work we contribute a large-scale classical benchmark dataset for the Lipkin–Meshkov–Glick (LMG) model generated using the Density Matrix Renormalization Group (DMRG). Using the Perlmutter supercomputer at NERSC \cite{nersc_perlmutter}, we obtain the full LMG ground-state energies for systems of up to $N=1400$ on the $V=1, W=0, \epsilon=1$ parameter set, with additional profiling simulations extending to $N=1550$. We additionally aggregate a phase diagram of ground-state energies for $N=[1,100]$ and $0.1$ increments from $V=[0,1]$. These datasets provide a high-precision classical reference across a wide range of system sizes and interaction strengths that can be leveraged for future benchmarking studies. We demonstrate the utility of the datasets by benchmarking the Variational Quantum Eigensolver (VQE) and Sample-based Quantum Diagonalization (SQD) algorithms on an IBM quantum processor. We demonstrate that these methods can reach significantly larger system sizes than previously reported for the LMG model. Our results establish performance baselines for each technique and offer insights into the computational trade-offs involved in simulating many-body systems at scale.

This paper is structured as follows. Section~\ref{sec:lipkin} presents a detailed review of the LMG model and its connection to phase diagrams. Section~\ref{sec:methods} outlines the computational methods, DMRG, VQE, and SQD, as applied to the LMG model. Section \ref{sec:results} presents the results of large-scale simulations performed using HPC and NISQ resources, with conclusions drawn in Section \ref{sec:conclusion}.

\section{The Lipkin Model}
\label{sec:lipkin}

The Lipkin-Meshkov-Glick (LMG) model~\cite{lipkin_validity_1965} has garnered significant attention as a benchmark for quantum simulations because it describes a many-body system that, while classically tractable, exhibits rich physical phenomena such as quantum phase transitions~\cite{castanos_classical_2006, leyvraz_large-n_2005, ribeiro_exact_2008}. Its tractability allows straightforward validation and verification of quantum results at scales typically beyond the reach of exact classical diagonalization. Previous work on simulating the LMG model on quantum devices has largely been confined to modest system sizes due to the constraints of current quantum hardware~\cite{cervia_lipkin_2021, Hlatshwayo_simulating_excited_states, robbins_benchmarking_2021, 
romero_solving_2022, hobday_variance_2024}.

The LMG model was originally introduced as an exactly solvable testbed to evaluate the validity of many-body approximation methods and theoretical frameworks \cite{lipkin_validity_1965}. It describes a system of \(N\) fermions distributed across two \(N\)-fold degenerate energy levels, interacting via collective, pairwise couplings that allow particles to scatter between these levels. Its Hamiltonian is given by:

\begin{align}
    H_N(\epsilon, V, W) = \frac{\epsilon}{2}&\sum_{p,\sigma}\sigma a^{\dag}_{p,\sigma} a_{p,\sigma} \nonumber \\
    +\frac{V}{2}&\sum_{p,p',\sigma}a^{\dag}_{p,\sigma}a^{\dag}_{p',\sigma} a_{p',-\sigma} a_{p,-\sigma}\nonumber \\
    +\frac{W}{2}&\sum_{p,p',\sigma}a^{\dag}_{p,\sigma}a^{\dag}_{p',-\sigma} a_{p',\sigma} a_{p,-\sigma}.
\end{align}

Here, the two energy levels are separated by a gap \(\epsilon\) and are labeled by the quantum number \(\sigma \in \{+1,-1\}\), whose sign corresponds to the upper and lower levels, respectively. Moreover, an additional quantum number \(p\) indexes the \(N\)-fold degenerate states within each level. The parameters \(V\) and \(W\) control the strength of the pairwise and exchange interactions, respectively. The former scatters a pair of particles from one level to the other, whereas the latter promotes one particle and demotes another. 

Given that each particle is restricted to one of two possible states, the LMG model naturally admits a quasi-spin formulation through the collective operators
\begin{align*}
    J_+ &= \sum_p a_{p,+1}^{\dagger} a_{p,-1}, \\
    J_- &= \sum_p a_{p,-1}^{\dagger} a_{p,+1}, \\
    J_z &= \frac{1}{2} \sum_{p,\sigma} \sigma\, a_{p,\sigma}^\dagger a_{p,\sigma}.
\end{align*}
These are termed quasi-spin operators because, although they are constructed from fermionic creation and annihilation operators rather than representing the intrinsic spin of any individual particle, they satisfy the same SU(2) commutation relations as angular momentum operators and encode the collective spin-like dynamics of the entire system. In terms of these operators, the Hamiltonian may be expressed as
\begin{equation}
    H_{N}(\epsilon, V, W) = \epsilon J_z + \frac{V}{2}(J_+^2 + J_-^2) + \frac{W}{2}(J_+J_- + J_-J_+).
\end{equation}

Here we consider a common variation on the model that assumes the spin exchange term has zero coupling strength ($W = 0$) \cite{Hlatshwayo_simulating_excited_states,cervia_lipkin_2021,lipkin_validity_1965} denoted by

\begin{equation}
   H_N(\epsilon, V, W=0) = \hat{H}_N(\epsilon, V) = \epsilon J_z + \frac{V}{2}(J_+^2 + J_-^2).
    \label{eq:h_W0_quasi}
\end{equation}

Next we write the Hamiltonian in terms of Pauli matrices. Following \cite{cervia_lipkin_2021}, we decompose the total quasi-spin as $J=\sum^N_{p=1}{J_p}$ where each $J_p$ acts on a spin $\frac{1}{2}$-degree of freedom. Further substituting 

\begin{align*}
    J_+ &= \sum^N_{p=1}J^+_p = \sum^N_{p=1} \frac{(X_p+ iY_p)}{2}\\
    J_- &= \sum^N_{p=1}J^-_p = \sum^N_{p=1} \frac{(X_p- iY_p)}{2} \\
    J_z &= \sum^N_{p=1}J^z_p = \sum^N_{p=1} \frac{Z_p}{2}.
\end{align*}

we can rewrite the Hamiltonian as

\begin{align}
    \hat{H}_N(\epsilon, V) = &\frac{\epsilon}{2}\sum^N_{p=1} Z_p + \frac{V}{2}\sum^N_{q>p\geq1} (X_pX_q - Y_pY_q).
\end{align}

We solve this Hamiltonian using the algorithmic methods outlined in the following section assuming $\epsilon = 1$.  

\section{Algorithmic Methods}
\label{sec:methods}
In this section, we describe the classical and quantum algorithms used to benchmark the performance of NISQ-era devices. Further implementation and experimental details can be found in the Appendix.

\subsection{Density Matrix Renormalization Group}

Tensor network methods are well-established tools for simulating quantum many-body systems
~\cite{Orus2014TensorNetworks}. 
Among these methods, the Density Matrix Renormalization Group (DMRG) algorithm has proven effective for determining the ground-states of one-dimensional (1D) quantum systems \cite{White1992DMRG, White1993DMRG}. Because of its strong accuracy and favorable scaling for many quantum lattice models, DMRG has also become an important classical baseline for benchmarking quantum algorithms and hardware \cite{Schollwock2011DMRG}. DMRG is a classical variational algorithm minimizing the expectation value 

\begin{align}
    E_{min} = min_\psi \bra{\psi} \hat{H} \ket{\psi}
\end{align}

 to approximate the ground-state of a quantum system. The quantum state is represented as a Matrix Product State (MPS) and the Hamiltonian as a Matrix Product Operator (MPO)--- both constructed from local tensors.

The MPS takes the form 
\begin{equation}
    \ket{\psi} = \sum_{\{s_i\}} A_{1}^{s_1} A_{2}^{s_2} \dots A_{N}^{s_N} \ket{s_1 s_2 \dots s_N},
\end{equation}
where \(A_{i}^{s_i}\) are local tensors, and \(s_i\) denotes the physical basis states at each site. The MPO representation of the Hamiltonian is expressed as 
\begin{equation}
    \hat{H} = \sum_{\{s_i, s_i'\}} \left( \prod_{i=1}^{N} W_i^{s_i s_i'} \right)
    \ket{s_1 \dots s_N} \bra{s_1' \dots s_N'},
\end{equation}
where each \(W_i^{s_i s_i'}\) is a rank-4 tensor acting on the site \(i\), with physical input index \(s_i\), output index \(s_i'\), and internal (bond) indices, forming a linear operator that acts on the full Hilbert space.

The algorithm sweeps across the MPS, contracting two neighboring tensors at sites $i$ and $i+1$ to form a two-site local tensor \(\ket{\phi}\), which serves as the variational ansatz for local optimization. The tensors outside this two-site block are contracted sequentially with the MPO and the conjugate MPS, accumulating from the left and right boundaries to form the left and right environment tensors \(L^{[i]}\) and \(R^{[i+1]}\), respectively. These are rank-3 tensors --- not states --- whose indices connect the physical, MPS bond, and MPO bond dimensions at the boundary of the active block. Together, they encode the influence of the remainder of the system on the local optimization and define an effective Hamiltonian acting on \(\ket{\phi}\):

\begin{equation}
    \hat{H}_{\text{eff}}^{[i,i+1]} = L^{[i]} \cdot \hat{H}^{[i,i+1]} \cdot R^{[i+1]},
\end{equation}

\noindent where the contraction $(\cdot)$ runs over the shared bond and MPO indices connecting the environments to the two-site Hamiltonian $\hat{H}^{[i,i+1]}$.

The effective Hamiltonian is then applied to $\ket{\phi}$, where we minimize its eigenvalue using the Davidson algorithm \cite{tensorsnet-dmrg} to find its corresponding minimum eigenstate.

\begin{equation}
    \hat{H}_{\text{eff}}^{[i,i+1]} \ket{\phi} = E^{[i,i+1]} \ket{\phi}
\end{equation}

After optimization, the new eigenstate \(\ket{\phi'}\) is reshaped and decomposed via singular value decomposition (SVD) back into two updated site tensors, replacing the original tensors at sites $i$ and $i+1$. This sweeping procedure continues iteratively across the entire MPS until convergence of the total energy or a predetermined maximum number of sweeps is reached.

For software implementation details of DMRG, refer to Appendix~\ref{sec:appendix-dmrg-implementation}. 

\subsection{VQE} 
The Variational Quantum Eigensolver (VQE) is one of the most widely studied quantum algorithms believed to be suitable for NISQ hardware ~\cite{Peruzzo_2014}. VQE estimates the minimal expectation value of a Hermitian operator with respect to a parameterized quantum circuit. By offloading the optimization to a classical computer, VQE significantly reduces the circuit depth and gate complexity required compared to purely quantum algorithms, making it practical for studying systems like the LMG model on near-term quantum devices~\cite{cervia_lipkin_2021, robbins_benchmarking_2021, romero_solving_2022, Hlatshwayo_simulating_excited_states}.

The method relies on the variational principle, which guarantees that the expectation value of an operator over any trial state provides an upper bound on the true ground state energy $E_{min}$:

\begin{equation}
       E_{min} \leq \bra{\psi(\theta)} \hat{H} \ket{\psi(\theta)}.
\end{equation}

The parameterized trial state $U(\theta)$, known as the ansatz, is iteratively refined by a classical optimizer (defined in Appendix~\ref{sec:appendix-vqe-optimizer}) to minimize this expectation value, thereby approaching the ground state. 

We define several variational ansatze to explore this compressed subspace. These circuits are composed of multiple layers of $R_y$ and CNOT gates. Because both $R_y$ rotations and CNOT operations are represented by purely real matrices, the resulting parameterized state vector is guaranteed to maintain strictly real amplitudes. This makes them specifically suited to produce states consistent with the real symmetric form of the Hamiltonian. Further details are provided in Appendix~\ref{sec:appendix-vqe-ansatz}.

Due to the ubiquity of VQE in the literature, we included it here to compare with DMRG and SQD. However, applying this procedure to the LMG Hamiltonian presents challenges in scaling the ansatz, as the Hilbert space grows exponentially with particle number, making classical optimization difficult. To address this, we employ a Hamiltonian compression scheme introduced in \cite{Hlatshwayo_simulating_excited_states} which reduces the effective system size by exploiting key symmetries of the model when setting $W=0$. This allows for the use of smaller ansatze, making VQE simulations more tractable at larger particle numbers at the cost of reduced generality within the Hamiltonian. It is important to emphasize here that all subsequent VQE results apply exclusively to this W=0 regime and do not represent the general $(W \neq 0)$ model.

The compression scheme works by expressing the \(N\)-particle LMG Hamiltonian in the collective spin basis \(\ket{J,M}\), where \(J = N/2\) and \(M\) is the spin projection, \(M \in \{-J,-J+1,\dots,J\}\), giving a total dimension of \(D = 2J+1\). As the interaction term \(V\) excites particles in pairs, it couples states that differ by \(\Delta M = \pm 2\). When \(W=0\), these are the only couplings that remain, which allows the Hamiltonian to be written in a block diagonal form separating the even and odd \(M\), which contain at most \(d = J+1\) states. By assigning these states integer labels, the Hamiltonian can be written as

\begin{equation}
    H = \sum_{k=0}^{d-1} a_k \ket{k}\bra{k} 
    + \sum_{k=0}^{d-2} b_k \bigl( \ket{k}\bra{k+1} + \ket{k+1}\bra{k} \bigr),
\end{equation}

with coefficients  

\begin{equation}
\begin{aligned}
    a_k &= \epsilon M_k, \\
    b_k &= -\frac{V}{2}\,F_{+}(M_k),
\end{aligned}
\end{equation}

where $M_k \coloneqq 2k - J$. Rather than mapping these integer labels directly to the computational basis via their binary representation, we utilize the Gray encoding \cite{gray1953pulse}, in which binary strings are ordered such that any two adjacent entries differ by only a single bit, reducing entangling depth.

\subsection{Sample-based Quantum Diagonalization}

Sample-based Quantum Diagonalization (SQD) is a hybrid quantum-classical algorithm in which a quantum device generates a bitstring distribution that is then used to project and diagonalize a target Hamiltonian on a classical computer~\cite{robledo-moreno2024chemistry}. The central motivation is that a quantum processor can efficiently sample from distributions that reflect the physical structure of the system --- distributions that a purely classical sampler could not efficiently reproduce. The classical diagonalization step is therefore not independent of the quantum sampling, but rather enabled by it: the quantum device identifies the relevant low-energy subspace, and the classical computer performs exact diagonalization within that subspace. This division of labor allows SQD to leverage quantum resources for the exponentially hard sampling problem while keeping the diagonalization tractable classically.

In the instance of the LMG model, the structure of the Hamiltonian imposes strong constraints on the structure of its eigenstates \cite{lipkin_validity_1965}. Because particles transition between energy levels in pairs, all accessible states maintain constant parity in their Hamming weight. Moreover, the Hamiltonian coefficients $\epsilon$, $V$, and $W$ are state independent due to degeneracies among the states labeled by these indices. As a result, the Hamiltonian treats all computational basis states with identical Hamming weight equivalently. This structure suggests that the eigenstates are restricted to a symmetry-defined subspace, which can be naturally expressed as a superposition of Dicke states with fixed parity, more generally expressed as

\begin{equation}
        \ket{\psi_N} = \sum_{j=0}^{L} C_{k_j} \ket{D^N_{k_j}}
\end{equation}

\noindent where $k_j = 2j + (N \bmod 2)$  and $L = \left\lfloor \frac{N + 2}{2} \right\rfloor.$ 

By definition, a Dicke state is an equal superposition of all computational basis states of $N$ qubits with exactly \(k\) ones and \(N-k\) zeros. These states can be efficiently prepared as quantum circuits using the method presented in~\cite{rafael2024Dicke}. Rather than preparing the full eigenstate \(\ket{\psi_N}\) as a single coherent circuit, we instead use an ensemble of individual Dicke state circuits whose collective measurement outcomes approximate the distribution of the true eigenstate. This approach distributes the complexity of the eigenstate across multiple circuits, as each targets a distinct component of the symmetry-restricted subspace. Refer to Appendix~\ref{sec:sqd-dicke-states} for information on the Dicke state preparation and the restricted LMG subspace structure.

The resultant measurement distributions are then used to project a Hamiltonian into a subspace of the full Hilbert space spanned by the most relevant computational basis states. For ground state estimation, this corresponds to measurement distributions that elucidate important computational basis states within the true ground state. The resulting projection reduces the dimensionality of the problem, enabling the classical diagonalization of Hamiltonian.
\begin{align}
\hat{H}^{S^{(k)}} &= \hat{P}^{S^{(k)}}\hat{H} \hat{P}^{S^{(k)}}, \nonumber \\\text{with} \quad \hat{P}^{S^{(k)}} &= \sum_{x \in S^{(k)}} \ket{x}\bra{x}.
\end{align}

\noindent where $|S^{(k)}|$, the number of retained bitstrings, controls the dimension of the projected Hamiltonian and thus the cost of the subsequent classical diagonalization. For ground state estimation, a good sample distribution concentrates on bitstrings with large overlap with the true ground state, so that diagonalization within $S^{(k)}$ closely approximates the exact result. In practice, $|S^{(k)}|$ is bounded by both the number of shots and the maximum subspace size that can be feasibly diagonalized classically. Here the maximum of 8192 shots places a strong limitation on the performance of SQD.

Measuring the ensemble of Dicke state circuits on a NISQ device inevitably induces some error due to noise \cite{cross_validating_2019, mind_the_gaps_2025}. However, because each circuit should yield bit strings with a fixed parity and Hamming weight, we are able to define a straightforward error mitigation scheme. We employ a two layer strategy: symmetry based post-selection followed by a modified version of the self-consistent configuration recovery strategy originally introduced in~\cite{robledo-moreno2024chemistry}. First, any bitstring whose measured Hamming weight, denoted \(w\), differs from the Dicke state's target value \(k\) is flagged as corrupted. Second, the bitstring is iteratively corrected by flipping a single bit, i.e., \(0\!\to\!1\) if \(w < k\), and \(1\!\to\!0\) if \(w > k\), until the corrected string reaches the target weight \(k\). Since all basis states of weight \(k\) are degenerate, we prioritize correcting each corrupted bitstring to a valid configuration that has not yet appeared in the sample, yielding a more complete representation of the Dicke state. This two step process ensures that the measurement data used for projection faithfully represents the symmetry-restricted subspace.

\section{Results}
\label{sec:results}
We describe our LMG benchmarking results of the classical and quantum approaches characterized in Section~\ref{sec:methods}. Algorithm solutions are considered successful only if they are within 1\% of the true ground state. Our objectives are to 1) establish a high-accuracy classical reference, 2) assess the performance of quantum-assisted algorithms relative to these references, and 3) characterize the scaling of quantum circuit resources required for these near-term implementations. 

\subsection{Classical Benchmark: DMRG}
First we present our DMRG dataset which serves as the reference against which LMG and SQD are compared. We evaluate DMRG for large-scale simulation on the Perlmutter supercomputer at the National Energy Research Scientific Computing Center (NERSC) \cite{nersc_perlmutter} to determine its long-term performance and scaling behavior.

\subsubsection{Comparison to Analytical Solutions}
\begin{figure}[htbp]
\centering
\includegraphics[width=\columnwidth]{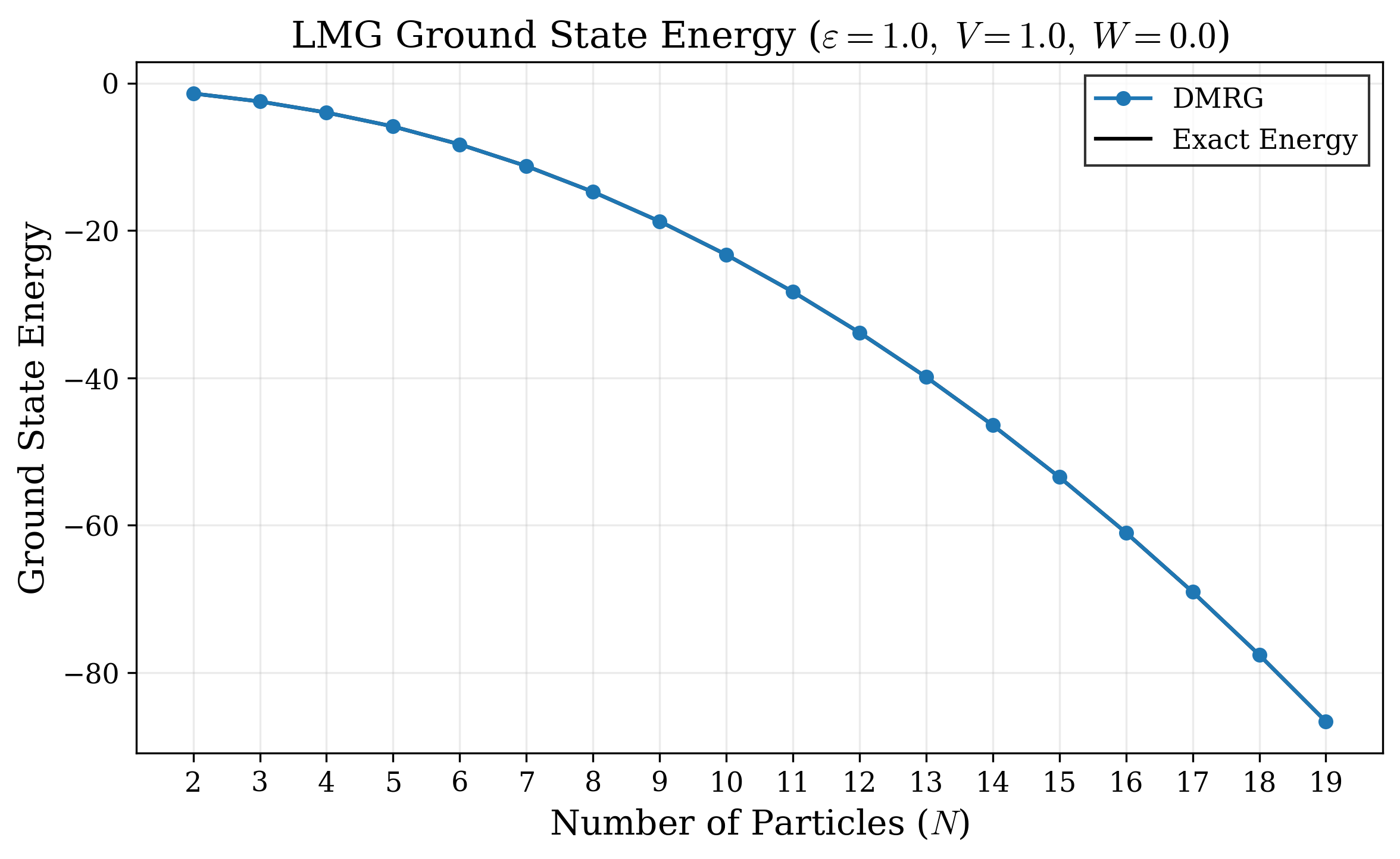}
\caption{Comparison between DMRG-computed ground-state energies and exact numerical solutions}
\label{fig:dmrg_stat_plot}
\end{figure}

\begin{figure}[htbp]
\centering
\includegraphics[width=\columnwidth]{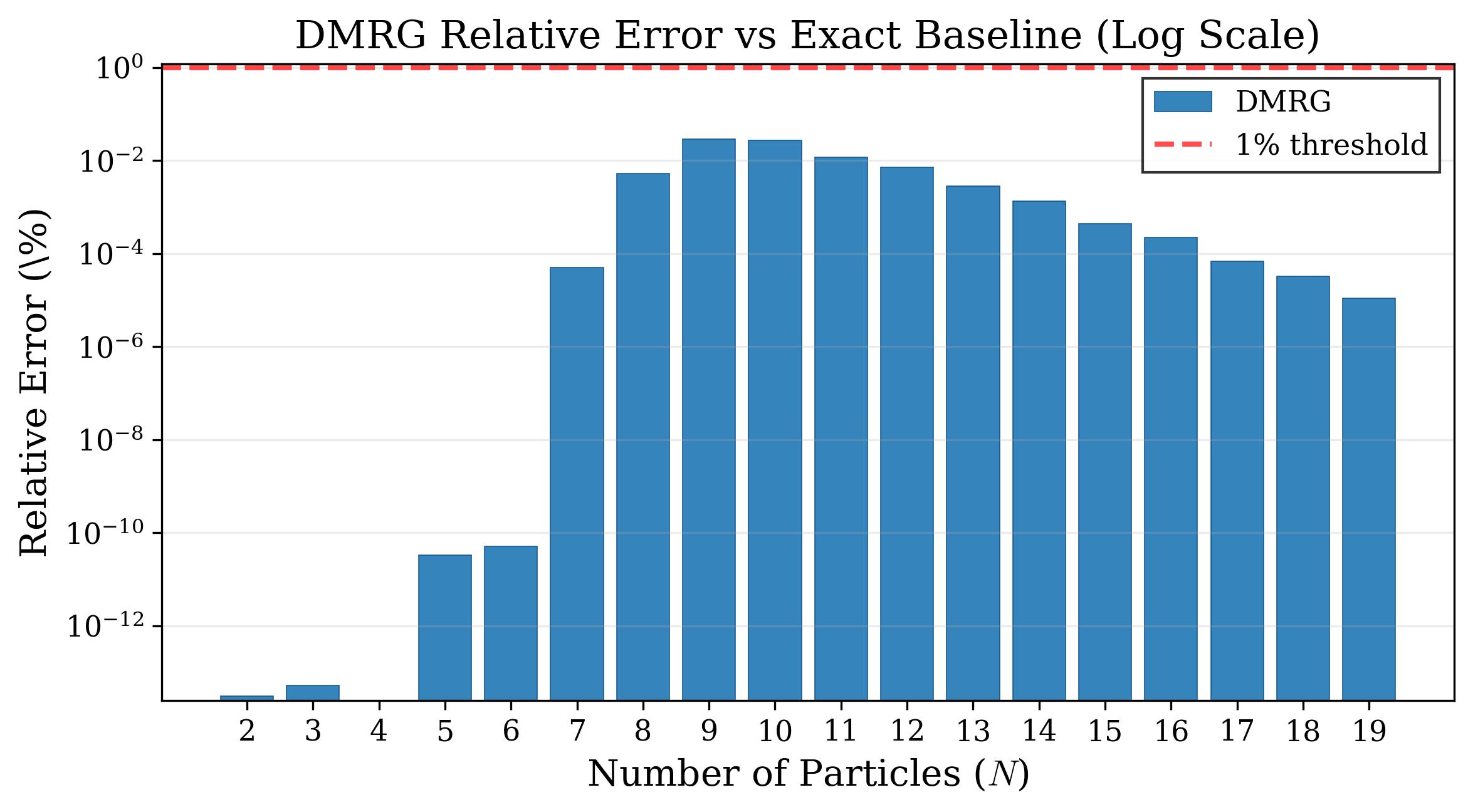}
\caption{Relative error of DMRG against exact numerical solutions on a logarithmic scale.}
\label{fig:dmrg_rel_error}
\end{figure}

We can investigate whether DMRG is a suitable (classical) candidate in solving for the LMG ground-state energy baselines. Figure~\ref{fig:dmrg_stat_plot} compares DMRG-computed ground-state energies alongside ground-state energies given by a direct diagonalization of the Hamiltonian for $N \leq 19$. Meanwhile, Figure~\ref{fig:dmrg_rel_error} illustrates the corresponding relative error plot. DMRG results report more than $99.99\%$ accuracy when compared to the exact LMG numerical solutions falling well within our $1\%$ error threshold. These results, along with the verification of correctness in published literature as well as statistical stability which are included in Appendix~\ref{sec:appendix-dmrg-validation}, indicate the DMRG results as a highly-reliable baseline.

\subsubsection{Ground-State Energies of Large-scale Systems}
Using the Perlmutter supercomputer at NERSC \cite{nersc_perlmutter}, we construct HPC-specific pipelines to determine LMG ground-state energies for large system sizes. These pipelines are expanded upon in Appendix~\ref{sec:appendix-dmrg-hpc}. 

\begin{figure}[htbp]
\centering
\includegraphics[width=\linewidth]{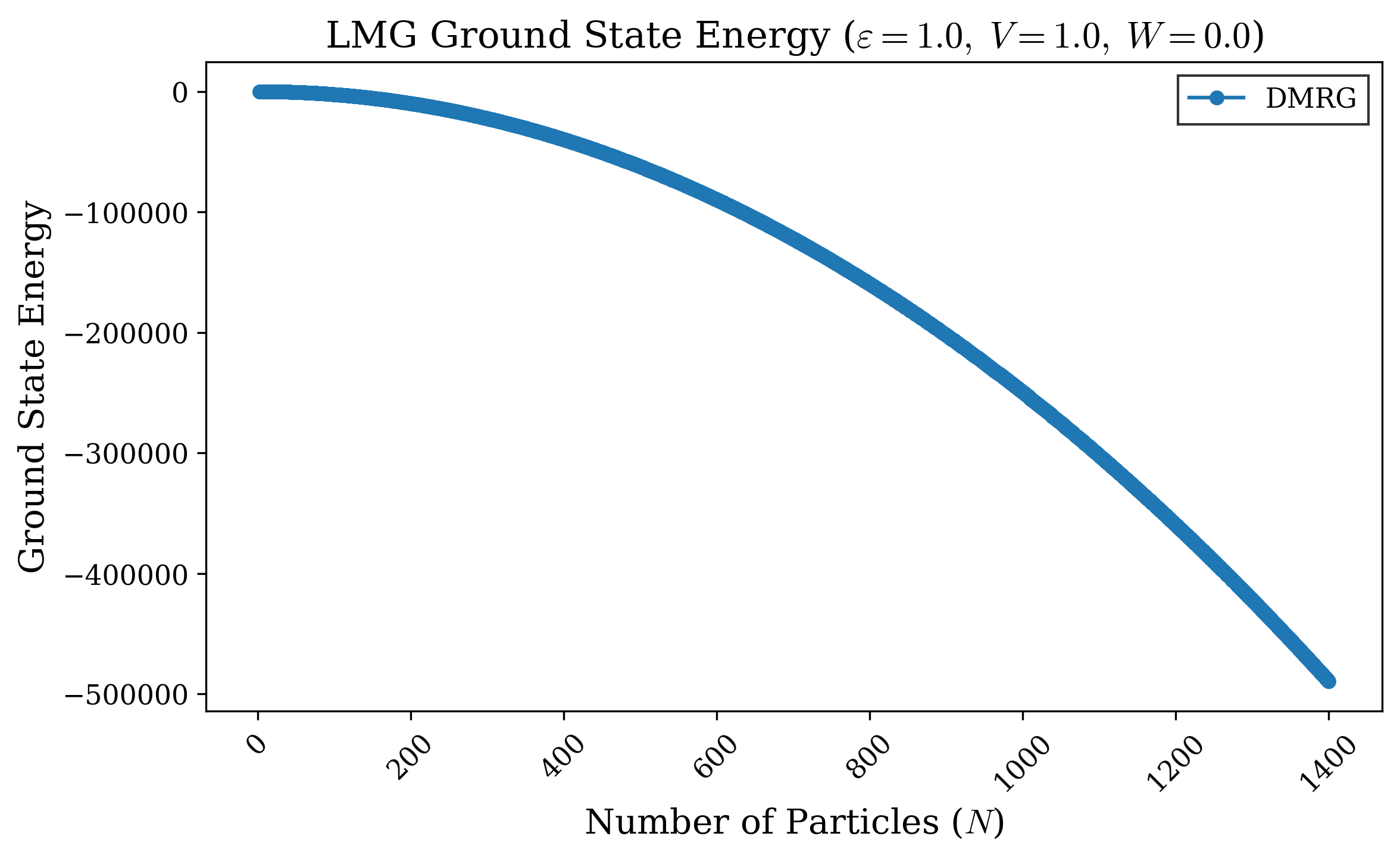}
\caption{Ground-state energy of the LMG model computed using DMRG as a function of particle number $N$.}
\label{fig:dmrg_energy_vs_n}
\end{figure}

We determine the dataset of ground-state energies of $\hat{H}_N(\epsilon = 1, V = 1, W = 0)$ for systems containing $N \in [2,1400]$ particles, plotted in Figure~\ref{fig:dmrg_energy_vs_n}.

As $N$ increases, the energy decreases monotonically indicating the collective interaction structure of the LMG Hamiltonian. The model illustrates a well-defined thermodynamic limit, so the smooth scaling behavior provides an important confirmation that the DMRG calculations converge reliably across the entire range of system sizes. This smooth energy curve affirms that the matrix product state representation is sufficient to capture the entanglement structure of the LMG ground state for large $N$.

\begin{figure}[htbp]
\centering
\includegraphics[width=\linewidth]{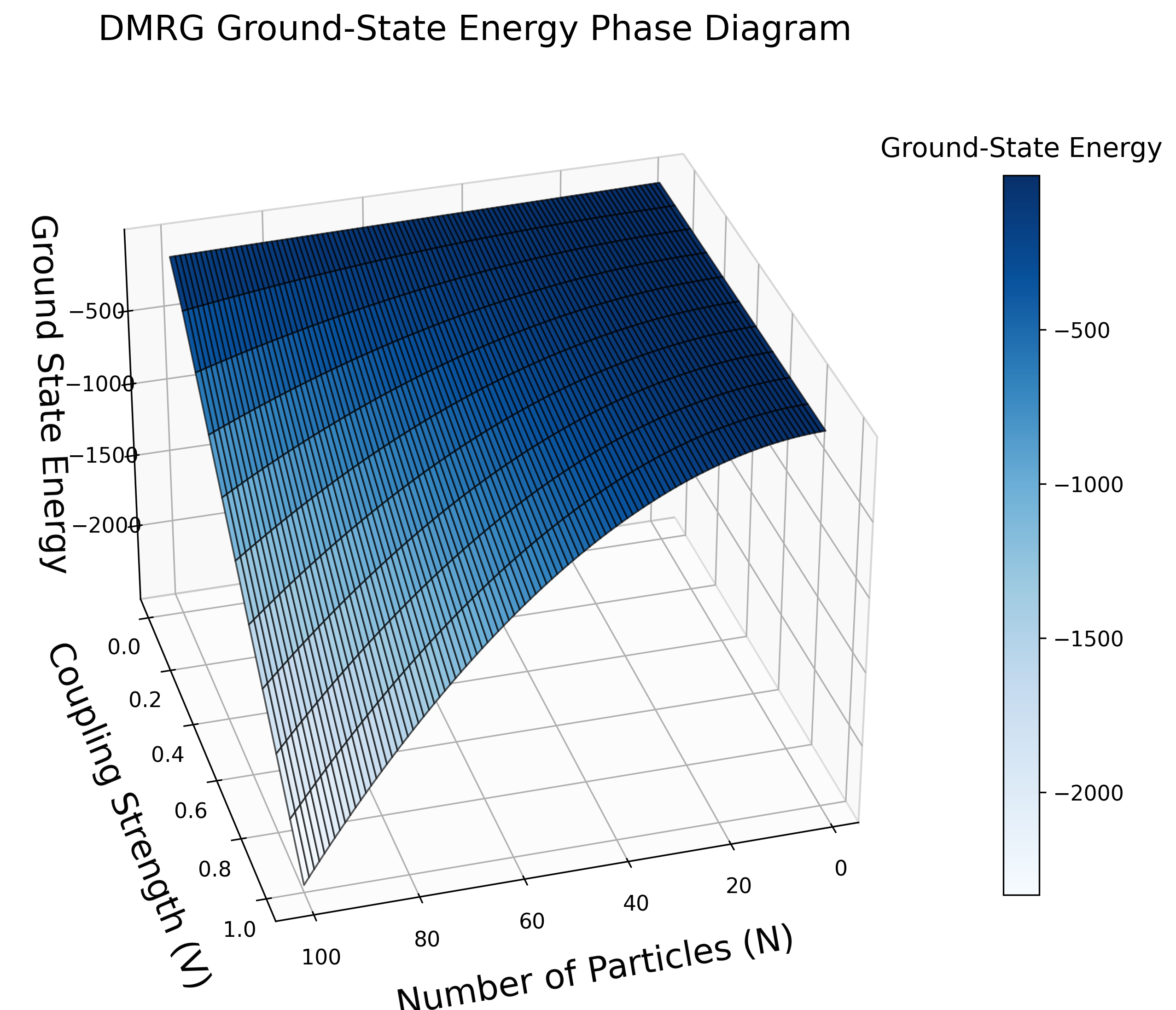}
\caption{DMRG-computed ground-state energy phase diagram as a function of particle number $N$ and interaction strength $V$.}
\label{fig:dmrg_phase_diagram}
\end{figure}

To further examine the behavior of the model across parameter space, we capture a two-dimensional scan of the ground-state energy over particle number $N \in [2, 100]$ and interaction strength $V \in [0, 1]$ in step sizes of 0.1 show in Figure~\ref{fig:dmrg_phase_diagram}. Similar to Figure~\ref{fig:dmrg_energy_vs_n}, the smooth plane of the energy landscape further confirms the numerical stability of the DMRG solver and the convergence of the tensor-network representation across the sampled region of the Hamiltonian parameter space. 

We also evaluate the computational cost of DMRG implementation, which becomes more evident as the system scales up. Appendix~\ref{sec:appendix-dmrg-profiling} details our profiling approach and findings for DMRG on different bond dimensions, system sizes, and hardware.

To our knowledge, the ground state energies displayed in Figure~\ref{fig:dmrg_energy_vs_n} and Figure~\ref{fig:dmrg_phase_diagram} are among the largest datasets recorded for the LMG model and are highly useful for benchmarking quantum algorithms. Access to the dataset is available in the Appendix. We demonstrate this utility by using these DMRG results as a reference benchmark for the VQE and SQD algorithms.


\subsection{VQE}

We apply VQE to the LMG model by obtaining results from an IBM device for $N \in [3, 25]$ with $V=1$. 

\begin{figure}[htbp]
    \centering
    \includegraphics[width=1.05\linewidth]{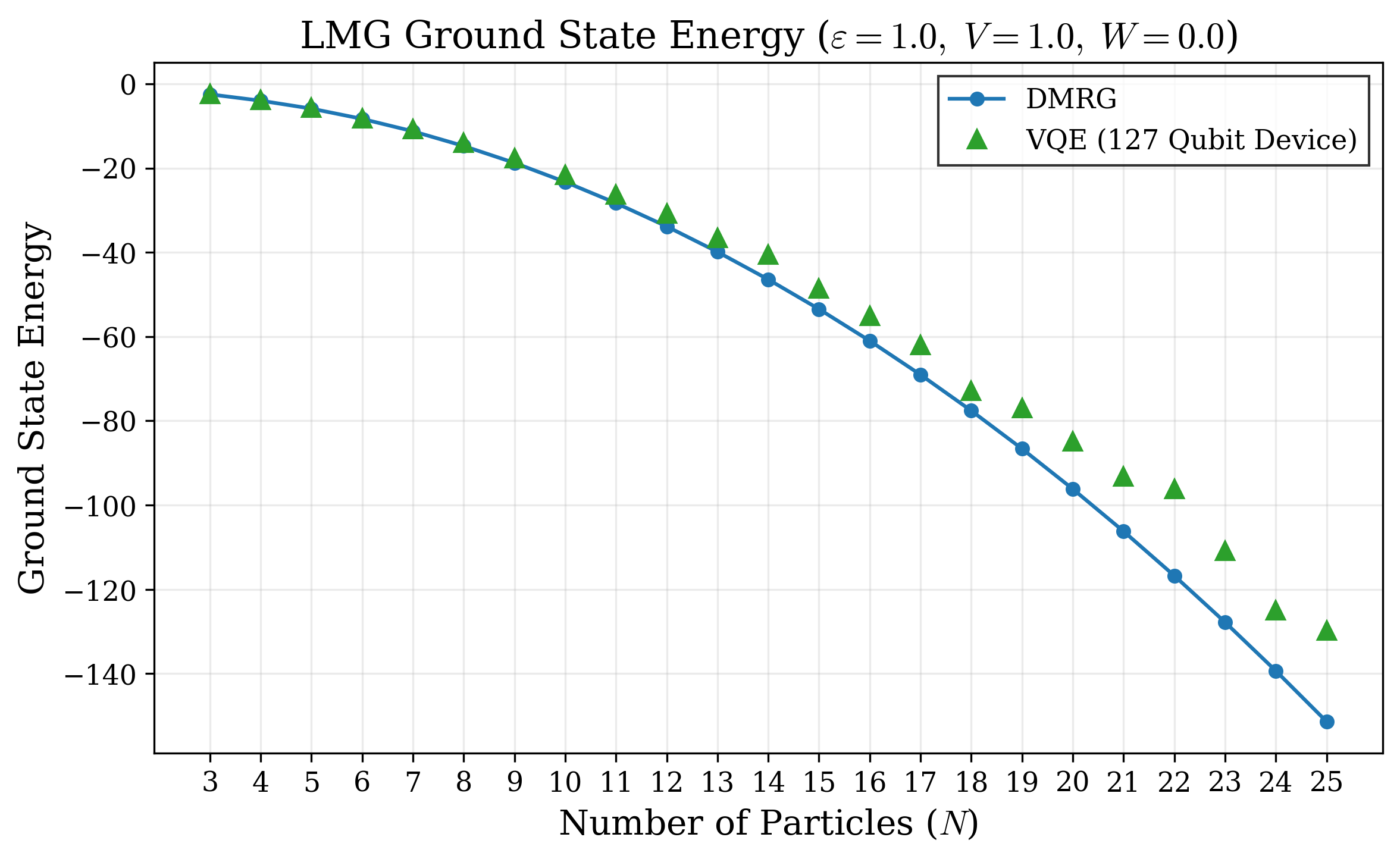}
    \caption{Iterations to convergence across four problem sizes: $N = 4,15,18,21$.}
    \label{fig:vqe_energy}
\end{figure}

Figure~\ref{fig:vqe_energy} displays the VQE ground-state energies in reference to the DMRG energies. From observation, while VQE tracks the DMRG/ground-truth curve well for small $N$, it becomes increasingly more unstable as $N$ grows, with the error gap widening steadily beyond $N \approx 15$ particles.

\begin{figure}[htbp]
    \centering
    \includegraphics[width=\linewidth]{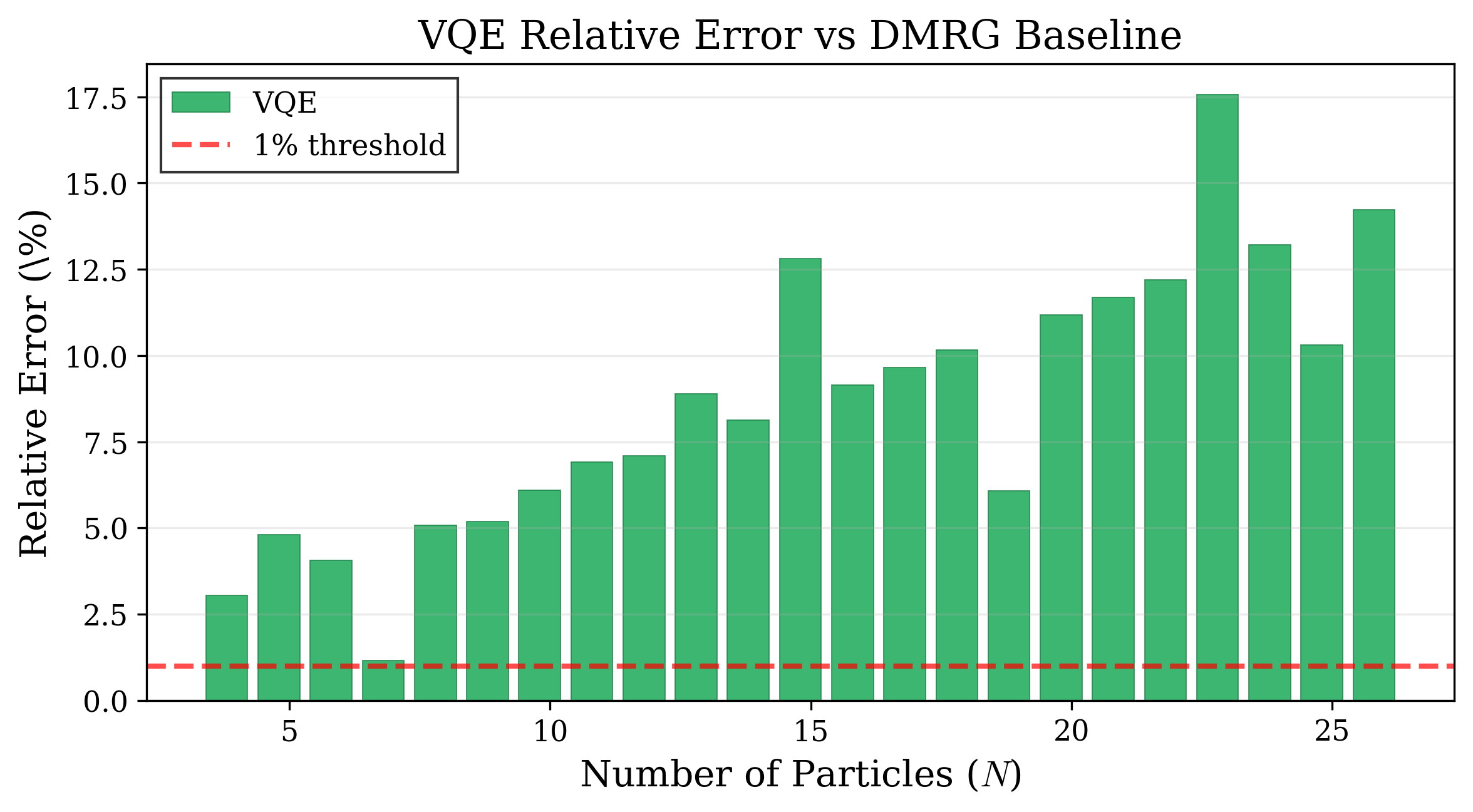}
    \caption{Relative error of VQE against DMRG baseline}
    \label{fig:vqe_error}
\end{figure}

Figure~\ref{fig:vqe_error} quantifies the relative error of VQE with respect to DMRG. Errors range from roughly $1\%$ to $5\%$ for small $N$, then climb into double digits beyond $N = 15$, reaching as high as ${\sim}17\%$ at $N = 23$. The growth is broadly monotonic but not uniform. There are local dips, such as at $N = 6$ and $N = 19$, likely reflecting favorable ansatz expressivity or optimizer convergence at particular problem sizes rather than a systematic trend. These results demonstrate that VQE cannot reliably achieve sub-$5\%$ accuracy beyond modest system sizes on current hardware. The errors reported here should be considered an optimistic lower bound, achievable only with careful hardware calibration and hyperparameter tuning.

\subsection{SQD}

We evaluate SQD on the LMG model using an ensemble of Dicke state circuits executed on the IBM processor. As described in Section~\ref{sec:methods}, each circuit targets a distinct symmetry sector of the LMG eigenstate, and their collective measurement outcomes are used to project and diagonalize the Hamiltonian classically.

\subsubsection{Accuracy and Shot-Budget Scaling}

\begin{figure}[htbp]
    \centering
    \includegraphics[width=\linewidth]{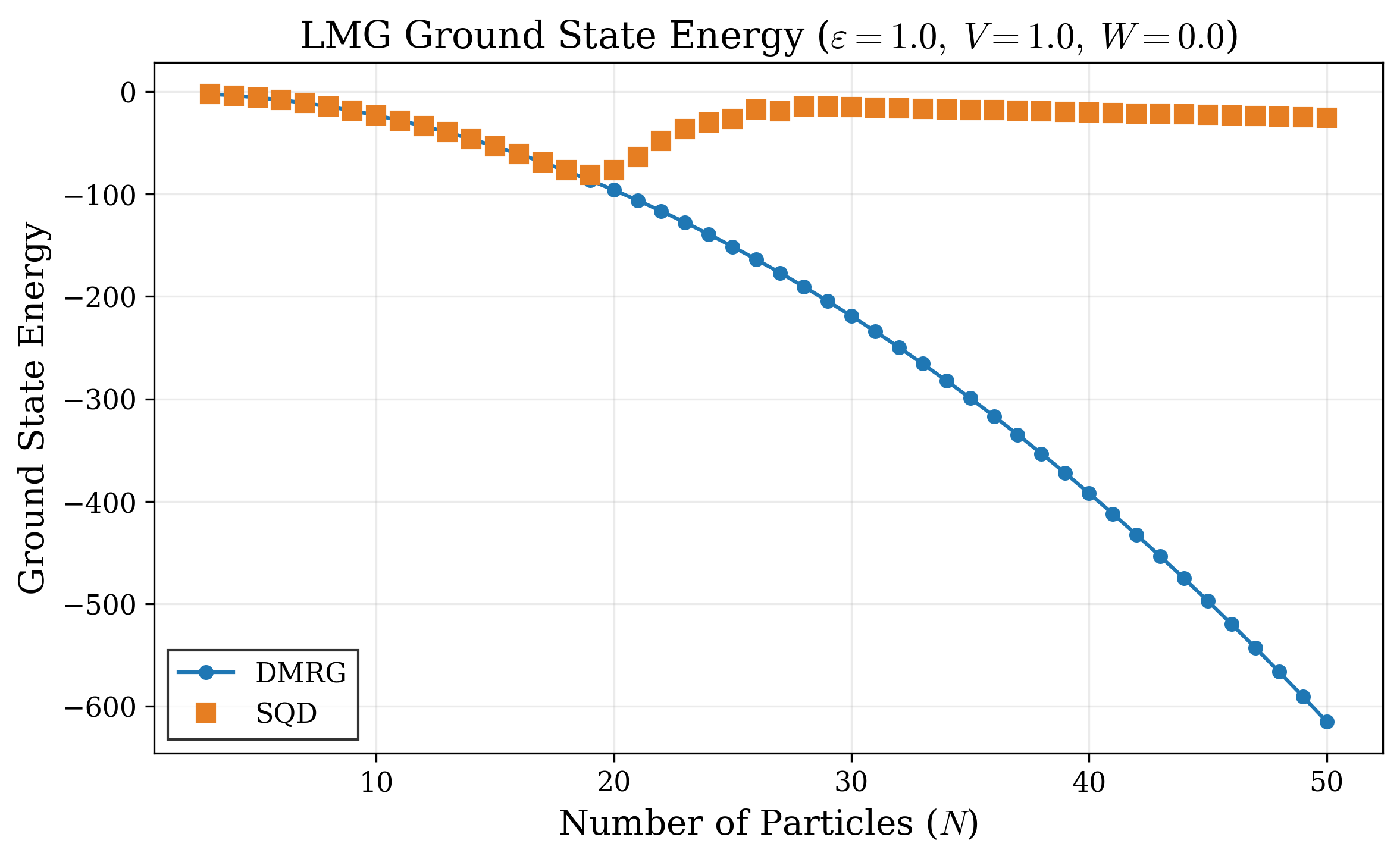}
    \caption{Ground-state energies at $V=1$, $W=0$ as a function of $N$, comparing DMRG (blue circles) and SQD on the IBM processor (red squares).}
    \label{fig:scatter}
\end{figure}

Figure~\ref{fig:scatter} compares SQD and DMRG ground-state energies at $V = 1$, $W = 0$ as a function of system size $N$, with SQD results obtained on the IBM processor using $10^6$ total shots distributed across the Dicke state ensemble. Two distinct regimes are visible. For $N < 20$,  SQD and DMRG agree to within ${<}0.5\%$, demonstrating near-exact diagonalization when the symmetry sector is sufficiently small. Beyond this threshold, SQD energies plateau and diverge substantially from the DMRG baseline, with the deviation growing rapidly with $N$.

The origin of this transition is the exponential growth of the LMG symmetry sector relative to the fixed shot budget. The total number of computational basis states within the relevant symmetry sector is $2^{N-1}$, while the processor supports a maximum of $10^6$ shots per job. When $2^{N-1} \ll 10^6$, repeated measurements collectively cover the full set of Dicke basis configurations, and projection onto the sampled subspace $S^{(k)}$ constitutes an essentially exact diagonalization. Once $2^{N-1}$ exceeds the shot budget, measurements can only sparsely sample the symmetry sector, so the projected subspace omits a growing fraction of energetically relevant basis states, and the diagonalized energy increasingly overestimates the true ground-state energy. This transition is consistent with the SQD circuit and sampling resource scaling derived in Appendix~\ref{sec:sqd-circuit-scaling}.

Figure~\ref{fig:error} quantifies this behavior through the relative error with respect to DMRG. For $N \leq 18$, the error remains below $1\%$, confirming that SQD reliably resolves the ground state in this regime. The error then rises sharply beyond $N \approx 20$, reaching ${\sim}95\%$ at $N = 50$ --- a rapid degradation that reflects the exponential growth of the symmetry-sector dimension relative to the fixed measurement budget, and establishes a clear practical ceiling for SQD on current hardware without algorithmic mitigation.

\begin{figure}[htbp]
    \centering
    \includegraphics[width=\linewidth]{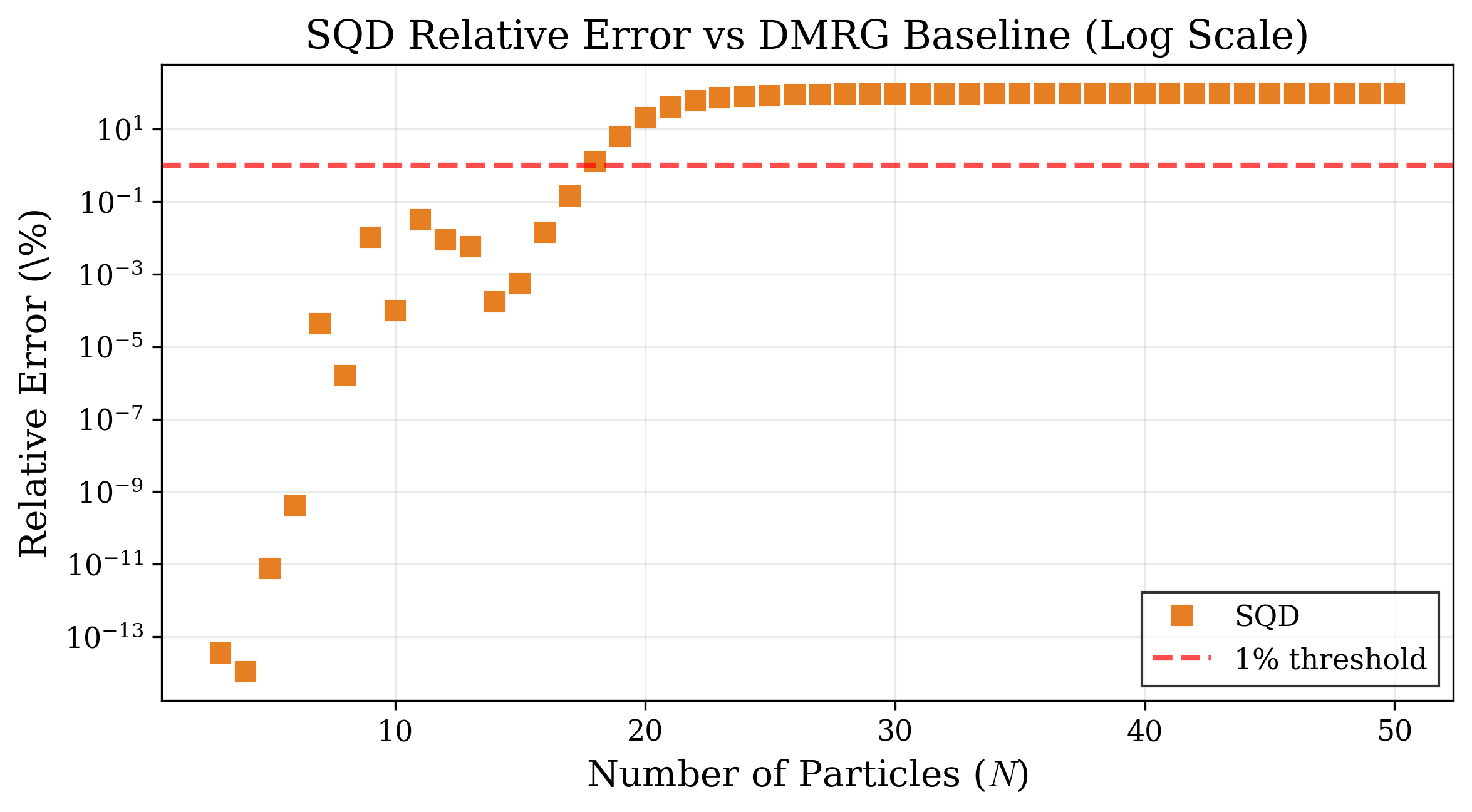}
    \caption{Relative error in SQD ground-state energy with respect to DMRG as a function of system size $N$.}
    \label{fig:error}
\end{figure}

To partially mitigate the shot-budget bottleneck, we implemented a weighted shot distribution that prioritizes dense Dicke states near $k \approx N/2$, which contribute the most basis states per circuit and thus contain the largest share of energetically relevant configurations. This heuristic improves subspace coverage at a fixed total shot count, though it cannot overcome the fundamental exponential scaling of the symmetry sector.

\subsubsection{Phase Diagram and Parameter Space Behavior}

The analysis above characterizes SQD performance along a single parameter line ($V=1$, $W=0$). To understand how accuracy and energy scale across the broader parameter space, we now examine the SQD energy and relative error over a two-dimensional grid of $(N, V)$ values.

\begin{figure}[htbp]
    \centering
    \hspace*{0\linewidth}
    \includegraphics[width=1.05\linewidth]{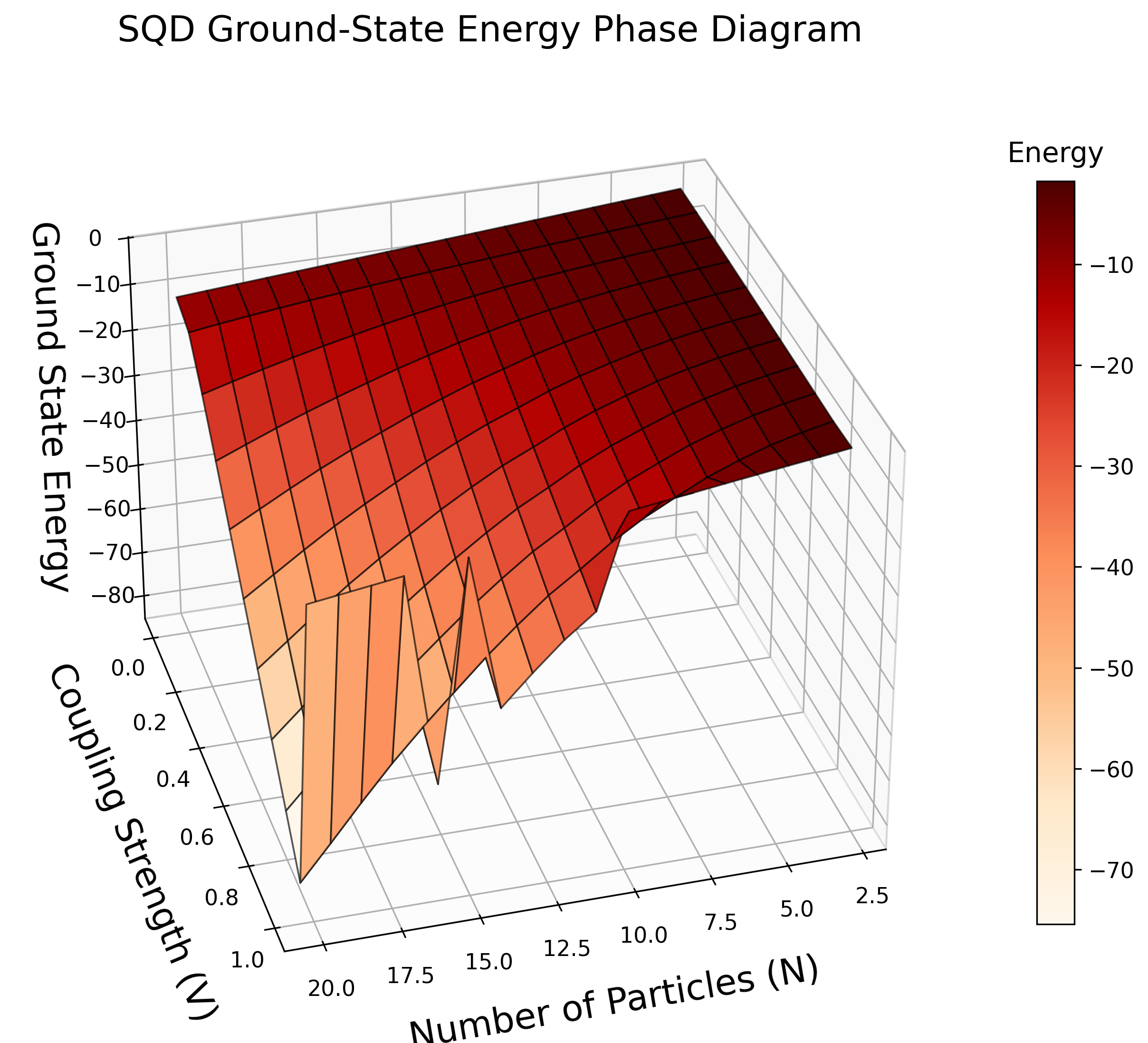}
    \caption{Phase diagram for SQD plotting the ground-state energy as a function of particle number $N$ and coupling strength $V$.}
    \label{fig:energy-surface-sqd}
\end{figure}

Figure~\ref{fig:energy-surface-sqd} shows the SQD phase diagram. The energy surface decreases monotonically with both $N$ and $V$, consistent with the DMRG phase diagram in Figure~\ref{fig:dmrg_phase_diagram}. The range of the SQD scan is substantially more limited in $N$ than the DMRG scan, reflecting the same shot-budget constraint identified above. 

\begin{figure}[htbp]
    \centering
    \hspace*{0\linewidth}
    \includegraphics[width=1.05\linewidth]{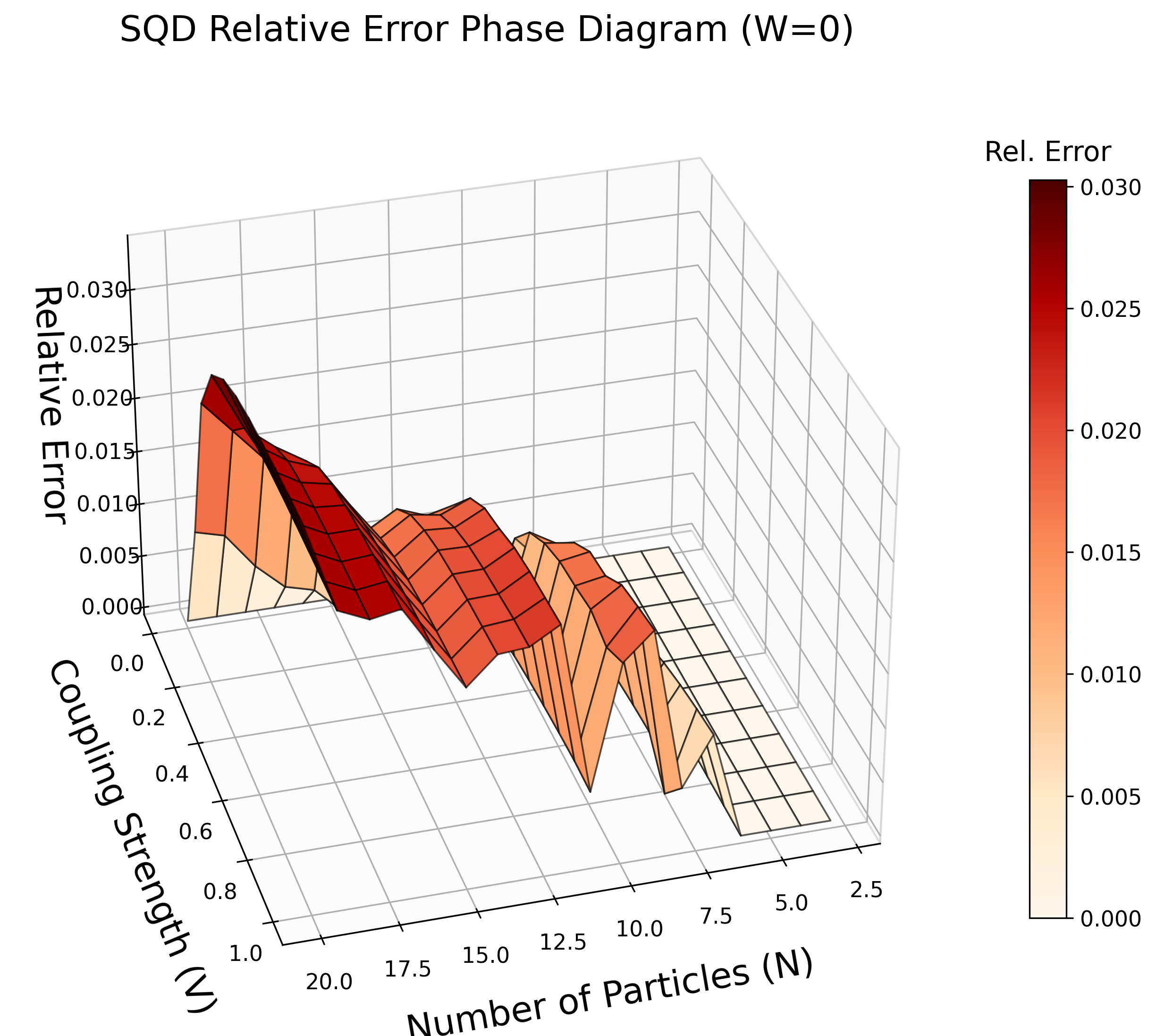}
    \caption{Phase diagram for plotting SQD energy error relative to true energy values as a function of particle number $N$ and coupling strength $V$.}
    \label{fig:energy-surface-sqd-rel-err}
\end{figure}

The relative error phase diagram in Figure~\ref{fig:energy-surface-sqd-rel-err} generalizes the $V=1$ result across the full parameter space. At $V = 0$, where particles do not interact, SQD agrees with the DMRG reference to machine precision across all system sizes, as the ground state is a simple product state trivially captured by any sample distribution. Away from $V = 0$, the error surface shows a clear overall increase with $N$, consistent with the shot-budget analysis above. There are also visible local oscillations in the error as a function of particle number whose pattern is not strictly periodic and whose origin is currently unclear; we leave their characterization as an open question for future work.

\subsection{Combined Comparison}

\begin{figure}[htbp]
    \centering
    \includegraphics[width=\linewidth]{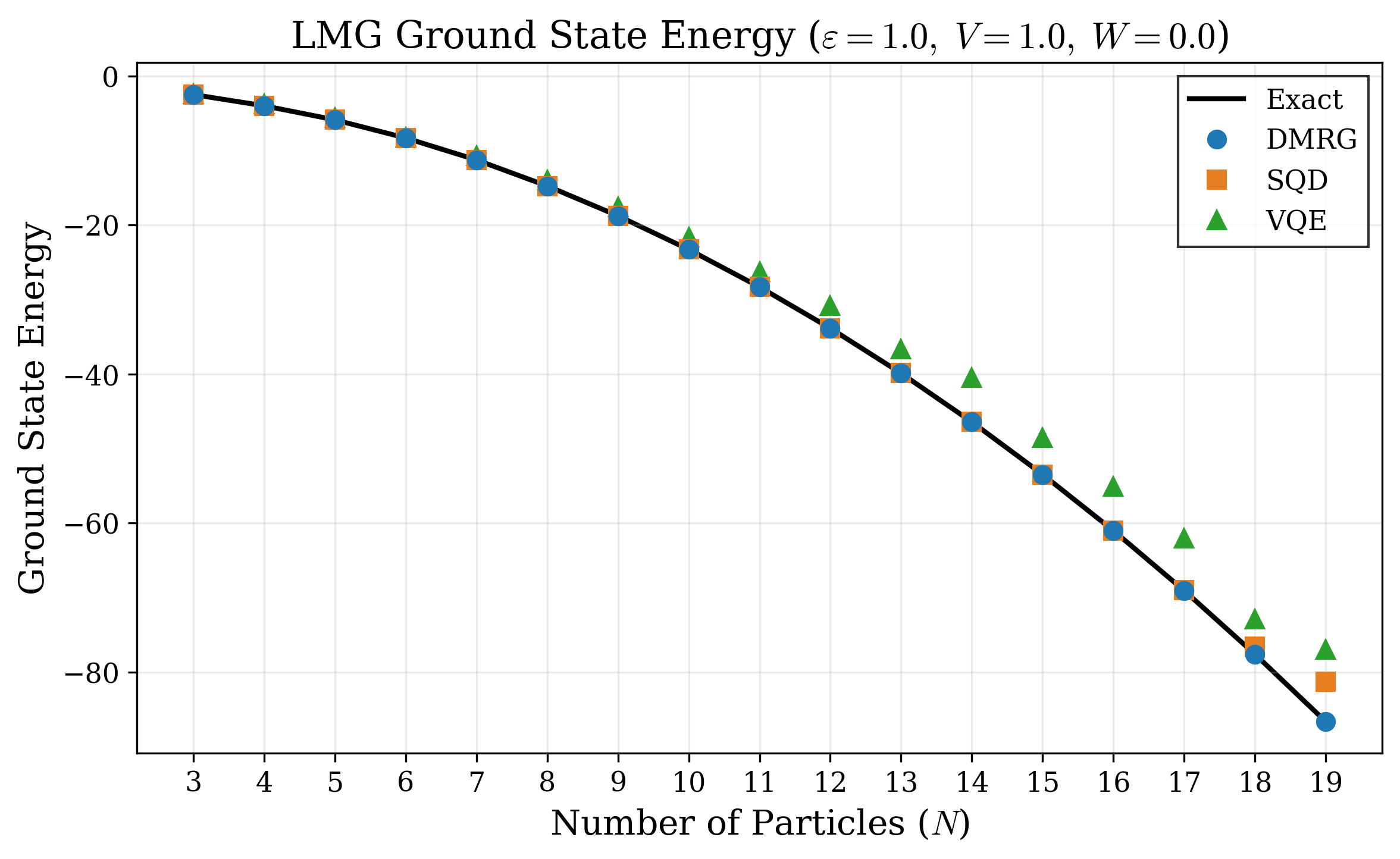}
    \caption{Ground-state energies computed by DMRG, SQD, and VQE compared against exact numerical diagonalization as a function of particle number $N$, at $\epsilon = 1.0$, $V = 1.0$, $W = 0.0$.}
    \label{fig:final_comparison}
\end{figure}

Figure~\ref{fig:final_comparison} brings together the results from all three methods alongside the exact numerical solution for $N \in [2, 19]$ at $\epsilon = 1.0$, $V = 1.0$, $W = 0.0$ --- the range where all three algorithms can be meaningfully compared. DMRG tracks the exact curve closely across the entire range, confirming its role as a reliable classical baseline. SQD matches the exact solution well for $N < 15$ before beginning to diverge, consistent with the shot-budget analysis in Section~\ref{sec:results}. VQE deviates visibly from the exact curve at all but the smallest system sizes, with errors that grow steadily with $N$.

\begin{figure}[htbp]
    \centering
    \includegraphics[width=\linewidth]{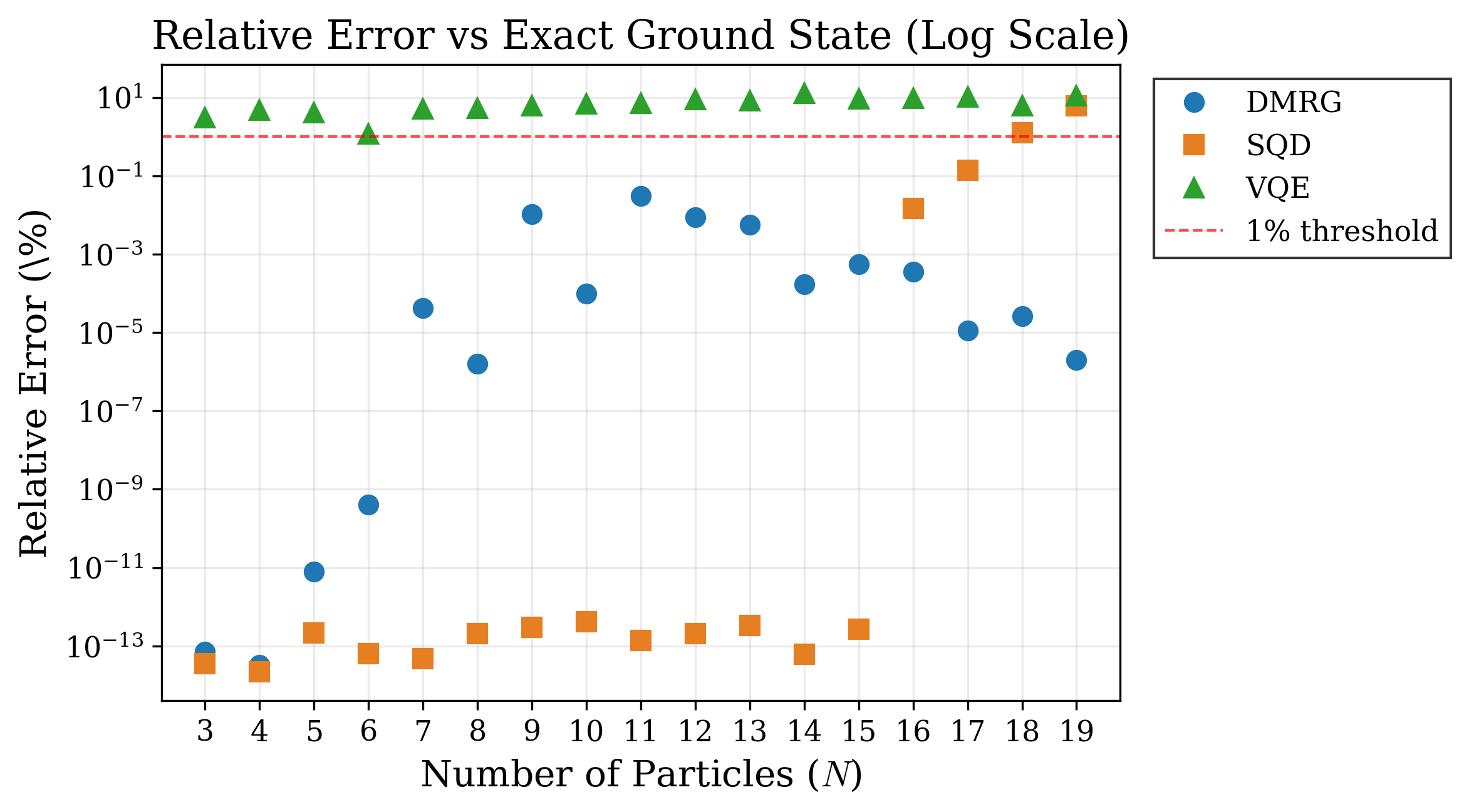}
    \caption{Relative error with respect to exact numerical diagonalization for DMRG, SQD, and VQE as a function of particle number $N$, shown on a log scale. The dashed red line marks the $1\%$ error threshold.}
    \label{fig:final_relative_error}
\end{figure}

Figure~\ref{fig:final_relative_error} shows the relative error of each method with respect to exact diagonalization on a log scale, which more clearly reveals the differences in accuracy across methods and system sizes. DMRG achieves the lowest errors overall, ranging from near machine precision at small $N$ to around $10^{-4}$\% at $N = 19$, reflecting the high fidelity of the tensor-network representation across this range. Interestingly, however, SQD outperforms DMRG in absolute accuracy for $N \in [3, 15]$, consistently sitting below $10^{-12}$\% in this regime --- a striking result that highlights the potential of quantum-assisted diagonalization when the symmetry sector is well covered by the available shot budget. This suggests that if the shot-budget limitation could be relieved, SQD would be an even more promising method, potentially surpassing DMRG in accuracy at small to moderate system sizes.

VQE errors range from approximately $1\%$ to $18\%$ across the full range of $N$, remaining consistently above the $1\%$ threshold and substantially less accurate than both DMRG and SQD throughout. Unlike SQD, which is highly accurate until it hits the shot-budget wall and then degrades sharply, VQE errors grow more gradually and steadily with $N$, reflecting the incremental difficulty of the classical optimization as the compressed Hilbert space grows. Both quantum methods therefore face a scaling challenge, but of qualitatively different character: SQD is highly accurate in its accessible regime but encounters a hard exponential ceiling, while VQE degrades more smoothly but never achieves the precision of either DMRG or SQD at these system sizes.

\section{Conclusion}
\label{sec:conclusion}

In this work, we used the DMRG algorithm to solve the LMG model, producing one of the largest datasets of LMG ground energies to extremely high accuracy. We then leveraged this dataset as a benchmark for the VQE and SQD algorithms.

Our primary classical contribution is twofold 
\begin{enumerate}
    \item A large-scale DMRG dataset of LMG ($V = 1$) ground-state energies spanning $N = 2$ to $N = 1400$ particles with profiling runs extending to $N = 1550$.
    \item An LMG phase diagram spanning $N \in [2, 100]$ with $V \in [0, 1]$ in increments of 0.1.
\end{enumerate}
To our knowledge, these represent the largest classical tensor-network simulations of the LMG model reported to date. The dataset is publicly accessible on GitHub \cite{QSL-NNL-P3} and provides a high-precision reference for benchmarking current and future quantum results. Our profiling analysis further revealed that Hamiltonian construction, not tensor contraction, dominates the total runtime at large $N$ due to the $\mathcal{O}(N^2)$ all-to-all interaction structure of the LMG Hamiltonian, limiting the practical benefit of GPU acceleration in our implementation.

On the quantum side, SQD on the IBM Heron processor achieved ground-state energies within ${<}0.5\%$ of the DMRG baseline for $N \leq 20$, demonstrating that quantum-assisted diagonalization can reach near-exact accuracy when the symmetry sector is fully coverable by the available shot budget. Beyond $N \approx 20$, where $2^{N-1}$ exceeds the $10^6$-shot limit of the processor, accuracy degrades sharply, reaching ${\sim}95\%$ relative error at $N = 50$. Circuit resource costs scale as $\mathcal{O}(N^2)$ in 2Q gate count, with maximum depths of ${\sim}1700$ at $N = 50$, approaching the coherence limits of current hardware.

Our VQE results show that errors range from ${\sim}1\%$ to $5\%$ at small system sizes and climb into double digits beyond $N \approx 15$, reaching as high as ${\sim}17\%$ at $N = 23$. VQE does not reliably achieve sub-$5\%$ accuracy beyond modest system sizes on current hardware, and optimizer convergence becomes increasingly erratic at larger $N$. Improved ansatz design or problem-specific initialization strategies will likely be needed before VQE can be considered competitive with classical methods on this model. Together, these results highlight the strengths and limitations of each approach. DMRG provides high-accuracy classical baselines at large scales but is ultimately bounded by the cost of MPO construction for all-to-all Hamiltonians. VQE, in its current form, struggles to scale beyond modest system sizes even with compression. SQD shows the most promise among the quantum methods, achieving competitive accuracy in the accessible regime, but its scalability is presently gated by shot budgets and circuit depth.

Future work should explore improved subspace sampling strategies for SQD to extend the accuracy regime beyond the shot-budget threshold, as well as alternative MPO construction methods to alleviate the Hamiltonian construction bottleneck in DMRG. Additionally, leveraging DMRG and HPC resources like Perlmutter should be explored further as a means of benchmarking quantum applications. By producing benchmarking datasets similar to those created here, tensor networks have the potential to provide the ultimate benchmark of quantum computation. Quantum computers can not demonstrate quantum advantage on a given application if they are unable to surpass the capabilities of classical tensor methods when applied to the same problem Hamiltonian.

\section*{Author Contributions}
M.B., R.D., and J.W. conducted the technical implementation, data collection, and drafting of sections for DMRG, VQE, and SQD respectively. J.A., H.Z., V.S., and B.M. provided project oversight and administrative support. All authors were involved in drafting, editing and reviewing of the final manuscript. Large language models were used for code repository refactoring and plotting.

\section*{Acknowledgments}
This work was performed as part of WISER Solutions Launchpad program, with the first three authors listed as the WISER Research fellows. This research used resources of the National Energy Research Scientific Computing Center (NERSC), a Department of Energy Office of Science User Facility under Contract No. DE-AC02-05CH11231 using NERSC award DDR-ERCAP0036702.  The authors additionally acknowledge the support provided by the IBM Quantum Enablement Team, and the use of IBM Quantum hardware resources.

This manuscript has been coauthored by the Naval Nuclear Laboratory, operated by Fluor Marine Propulsion, LLC under contract No. 89233018CNR000004 with the U.S. Department of Energy. The United States Government retains and the publisher, by accepting the article for publication, acknowledges that the United States Government retains a non-exclusive, paid-up, irrevocable, world-wide license to publish, distribute, translate, duplicate, exhibit, and perform the published form of this manuscript, or allow others to do so, for United States Government purposes.

\printbibliography

\onecolumn\newpage
\appendix
\section{Appendix}
This appendix provides additional details on the implementations of each method described in the main body of the paper. Section~\ref{sec:appendix-dmrg} expands on our implementation of the DMRG method, including details on the software packages used, validation of the method's accuracy, the HPC environment used for our runs on the Perlmutter supercomputer, and the method's computational cost and scaling. Section~\ref{sec:appendix-vqe} similarly describes implementation details of VQE, including the optimizer and ansatz used as well as training data. Finally, Section~\ref{sec:appendix-sqd} elaborates on our implementation of SQD method including Dicke state preparation and circuit cost scaling. The benchmarking code and analysis pipelines are available at: \url{https://github.com/Quantum-Solutions-Launchpad/LMG-Model-Quantum-Benchmarking-with-Tensor-Networks}. Our DMRG, SQD, and VQE datasets described in the paper are available upon request, please contact Vardaan Sahgal and the WISER team for more information. 


\section{DMRG Implementation}\label{sec:appendix-dmrg}
We document additional details of DMRG here, including our software implementation, algorithm verification, modifications made for use on HPC systems, and DMRG cost and scaling behaviors.

\subsection{Software Description}\label{sec:appendix-dmrg-implementation}

Our DMRG simulations are implemented in the Julia programming language (v1.11.7) with tensor-network libraries \texttt{ITensors.jl} and \texttt{ITensorMPS.jl} providing for representations of matrix product states (MPS) and matrix product operators (MPO) \cite{Fishman2020ITensor, ITensorsJulia}. The main computational kernel is constructed as the custom Julia package, \texttt{SingleDMRGLMG} \cite{SingleDMRGLMG}. 

This package provides the function $\texttt{run\_dmrg\_lmg\_single}(N,\epsilon,V,W)$. It builds the Hamiltonian, initializes the MPS, and performs the DMRG optimization. It also uses an optionally configurable sweep schedule consisting of the number of sweeps, a truncation cutoff, and a maximum bond dimension schedule. In this work we use fixed sweep schedules with gradually increasing bond dimensions to ensure convergence of the ground-state energy. By default, the solver uses the following DMRG sweep schedule:
\begin{verbatim}
nsweeps::Int = 5
maxdim::Vector{Int} = [10,20,100,100,200]
cutoff::Float64 = 1e-10
\end{verbatim}
These parameters were selected empirically (see ~\ref{sec:appendix-dmrg-validation} for details) to ensure convergence of the ground-state energy across the system sizes studied, while keeping contractions computationally feasible for large-scale simulations. We also compile the package into a custom Julia system image to eliminate just-in-time (JIT) compilation overhead during batch workloads. This significantly reduces startup latency per process, which is crucial when launching many parallel jobs.

\subsection{Validation of DMRG}\label{sec:appendix-dmrg-validation}
Due to the wide variety of alternative formulations of the LMG model, we first establish the accuracy of DMRG results by comparing our results to benchmarking data from \cite{lipkin_validity_1965, Hlatshwayo_simulating_excited_states}. 

\begin{figure}[!ht]
    \centering
  \includegraphics[width =0.5\columnwidth]{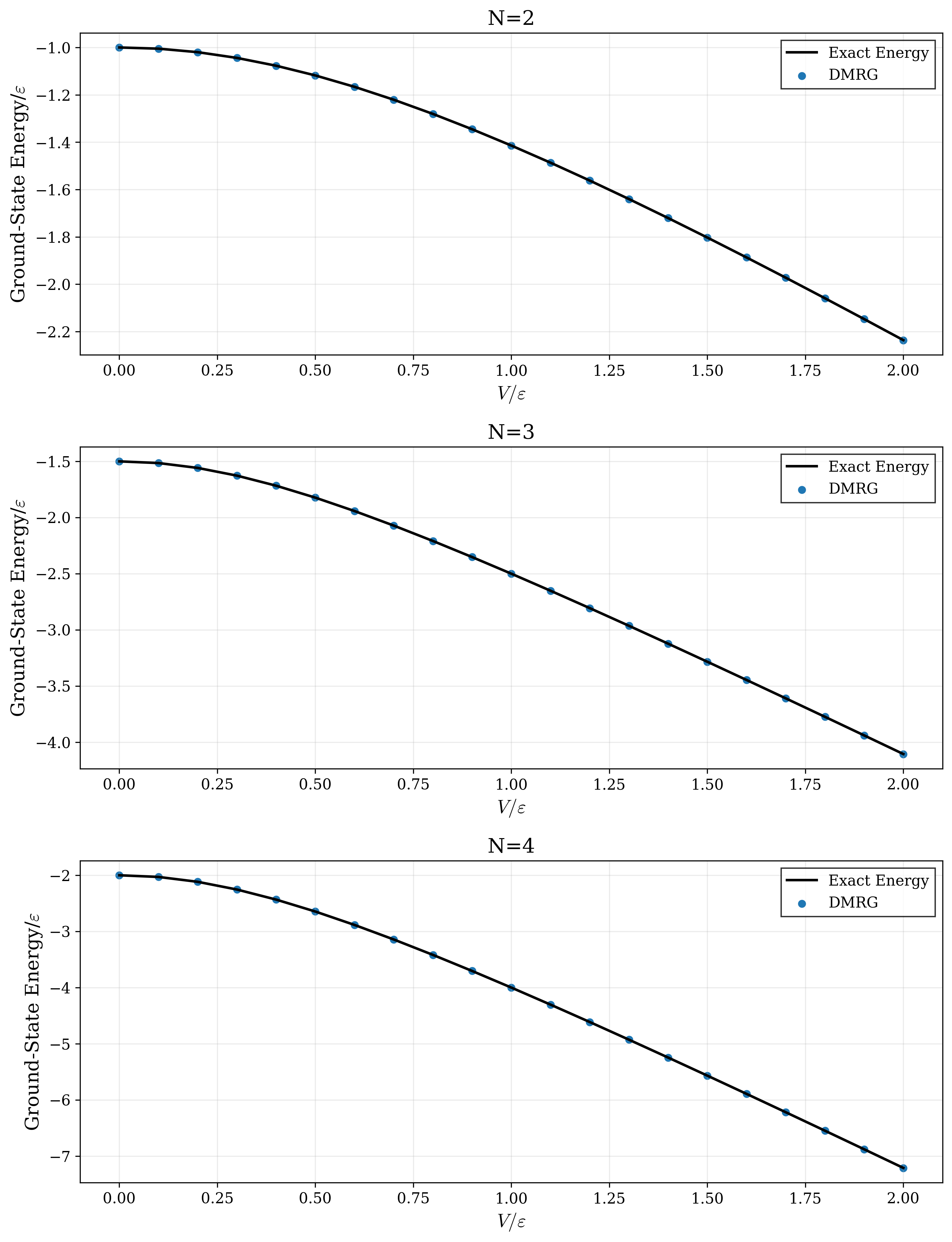}
\caption{Replication of benchmarking plots from Hlatshwayo et al. \cite{Hlatshwayo_simulating_excited_states} using DMRG simulations of the LMG model.}
\label{fig:Hlatshwayo}
\end{figure}

In Fig 4-6 of "Simulating excited states of the Lipkin model on a quantum computer", authors Hlatshwayo et al. produce benchmarking plots for the LMG Hamiltonian energy spectra. Replicating Hamiltonian and parameter settings described in ~\cite{Hlatshwayo_simulating_excited_states} --- but with DMRG --- we generate Figure~\ref{fig:Hlatshwayo}. Observe that the resulting curves produced by DMRG align with the qualitative and quantitative trends of the published results, confirming a correct implementation of the Hamiltonian construction and simulation pipeline. 

\begin{figure}[!ht]
  \centering
    \includegraphics[width=.5\textwidth]{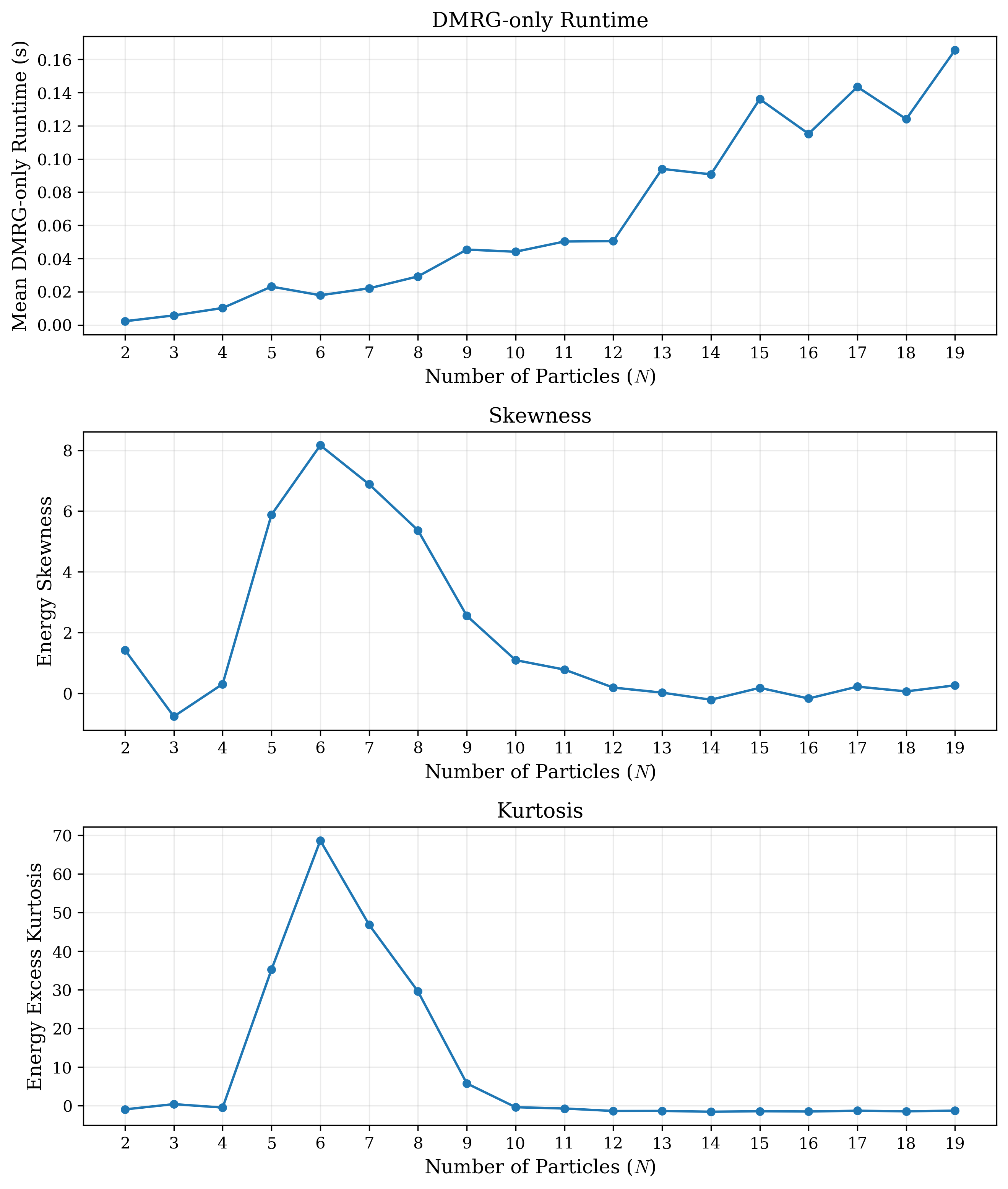}
    \caption{Statistical analysis of DMRG ground-state energy estimates across repeated runs ($X=100$) for $N=[2,19]$, with respect to the numerical diagonalization results for the LMG Hamiltonian (exact numerical solutions). Plots reveal runtime, skew, and kurtosis respectively.}
    \label{fig:statistical_example}
\end{figure}

Having validated against the literature, next we compare DMRG to the exact numerical solutions obtained from a classical eigenvalue solver. This classical solver determines the ground-state energy using a sparse eigensolver, and returns the minimum eigenvalue of the Hamiltonian matrix. Its results, limited only by compute of system sizes where the full Hamiltonian could still be diagonalized numerically, act as the ground-truth reference for the model for $H(\epsilon = 1, V = 1.0, W = 0.0)$ for $N\in[2,19]$. As DMRG is an iterative optimization procedure, repeated runs with identical parameters yield slight numerical variation depending on initialization and truncation effects. In Figure~\ref{fig:statistical_example} we perform $X = 100$ runs of DMRG for each N. 

\begin{figure}[!ht]
    \centering
    \includegraphics[width=.5\textwidth]{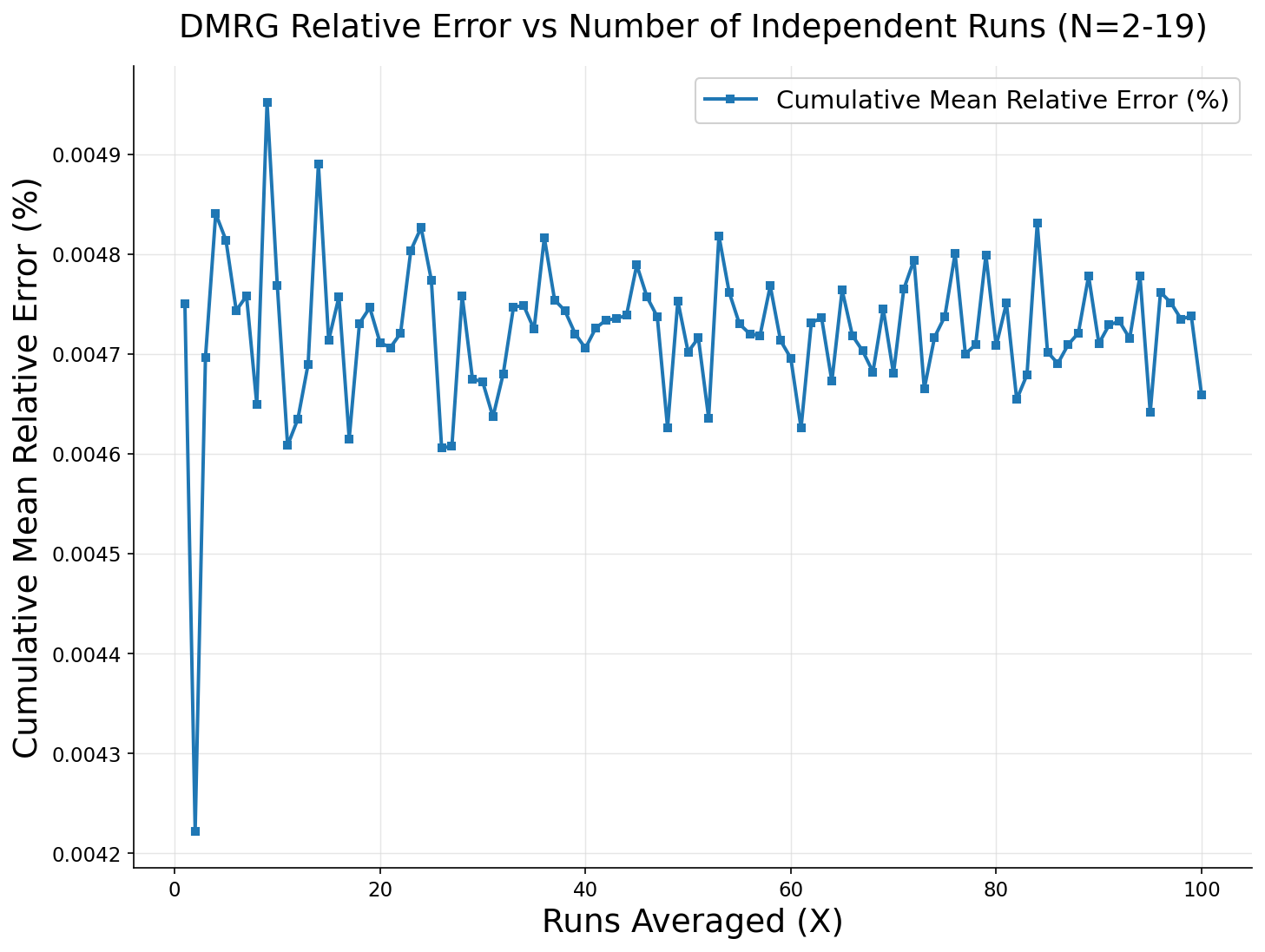}
    \caption{Cumulative mean relative error of DMRG ground-state energies as a function of the number of repeated runs $X$.}
    \label{fig:error_v_runs}
\end{figure}
    
Each run uses identical Hamiltonian parameters but different random initializations of the MPS. The resulting distributions are tightly concentrated around the exact energy values, with extremely small standard deviations and negligible skew. This indicates that the DMRG solver produces highly stable energy estimates across repeated runs despite differences in initialization and truncation effects. The mean DMRG energies align closely with the exact diagonalization results, demonstrating that the tensor-network representation accurately captures the ground-state structure of the LMG Hamiltonian in this regime.

Additionally, in Figure~\ref{fig:error_v_runs} we investigate how increasing the number of independent runs $X$ affects the estimated average error. As $X$ increases, the cumulative mean error rapidly stabilizes and converges to a constant value, indicating that a relatively small number of repeated runs is sufficient to obtain reliable estimates of the ground-state energy. This behavior further supports the numerical stability of the DMRG solver and demonstrates that statistical fluctuations from initialization do not significantly affect the final results. Having established the correctness and stability of the DMRG solver, we use DMRG as a reliable baseline for quantum algorithms.

\subsection{HPC Environment}\label{sec:appendix-dmrg-hpc}

GPU acceleration uses the \texttt{CUDA.jl} backend. When GPU execution is enabled on $\texttt{run\_dmrg\_lmg\_single}(N,\epsilon,V,W)$ in $\texttt{SingleDMRGLMG}$, the MPO and MPS are converted to GPU arrays, allowing tensor contractions in DMRG to be executed on CUDA kernels. Scalar indexing was disabled to prevent CPU fallback during the DMRG compute stage.

To safeguard jobs against out-of-memory errors for large-scale simulations, an additional pipeline was developed specifically for large jobs. This includes features such as memory monitoring and recovery mechanisms. The host and GPU memory usage are periodically checked and garbage collection and CUDA memory reclamation are triggered when necessary. Additionally, if a job unexpectedly fails, the program logs any existing data and proceeds with running DMRG-LMG on the next set of parameters. These safeguards allow long-running scans to complete even when individual runs encounter resource limits.

DMRG simulations were executed on the Perlmutter supercomputer at NERSC using the SLURM workload manager. We analyzed DMRG performance on both GPU and CPU nodes. Each GPU node contains four NVIDIA A100 GPUs, so we specify GPU jobs as follows:

\begin{verbatim}
--ntasks-per-node=4
--gpu-bind=single:1
\end{verbatim}

where one SLURM task is bound to one GPU, and there is one process per GPU. We use multi-threaded execution on CPU nodes. The number of ranks per node is dependent on the size of the job.

Our DMRG-LMG simulation pipeline runs a large number of parameter specifications across independent DMRG runs. We used a round-robin technique to select SLURM ranks with balanced workloads, since runtime grows nonlinearly with system size. Each rank/process outputs data to its own JSONL file. This prevents file-write contention during parallel execution.

A run produces a JSONL record containing

\begin{itemize}
\item system size $N$
\item Hamiltonian parameters $(\epsilon,V,W)$
\item computed ground-state energy
\item runtime
\item timestamp
\end{itemize}

Results are saved incrementally in batches to minimize I/O overhead and prevent data loss during long runs.

\subsection{Computational Cost \& Scaling} \label{sec:appendix-dmrg-profiling}

We developed a separate profiling utility that chunks $\texttt{run\_dmrg\_lmg\_single}(N,\epsilon,V,W)$ into its phases, collecting runtime data such as Hamiltonian construction time, MPS initialization time, GPU transfer time, and DMRG optimization time. This allows detailed analysis of the contributions of CPU pre-processing relative to GPU/CPU tensor contraction workloads.

The complete codebase used to generate the DMRG results, including job submission scripts and analysis utilities, is available in our public Github repository linked above. It contains the implementation of the LMG Hamiltonian construction, the DMRG execution pipeline, and scripts used to generate the datasets and figures presented in this work.

Table~\ref{profiling_table} summarizes representative runtimes required to compute ground-state energies across several system sizes. The table reports both total wall time and the time spent exclusively in the DMRG 
sweeping stage. As expected, the runtime increases with system size due to both the increasing cost of tensor contractions and the growing complexity of constructing the Hamiltonian MPO.

To evaluate the impact of hardware acceleration, Table~\ref{profiling_table} also compares the total wall runtime of the two pipelines. Hamiltonian construction dominates the total computational cost, accounting for more than $99.7\%$ of the wall time for large system sizes. Because this stage is CPU-bound in both implementations, GPU acceleration provides only limited overall speedup. This bottleneck arises from the all-to-all interaction structure of the LMG  Hamiltonian. Constructing the MPO requires incorporating $O(N^2)$ interaction terms, each contributing additional tensor contractions and index bookkeeping during initialization. As a result, the Hamiltonian construction stage scales more rapidly with system size than the subsequent DMRG sweeps. Even when GPU acceleration is applied to the tensor-network optimization stage, the overall runtime remains dominated by CPU pre-processing.

\begin{table}[t]
\centering
\footnotesize
\setlength{\tabcolsep}{4pt}
\begin{tabular}{c c c c c}
\hline
$N$ & CPU-wall & GPU-wall & CPU-DMRG & GPU-DMRG \\
\hline

\multicolumn{5}{c}{\textbf{maxdim: [10,20,100,100,200]}} \\
100  & 4.05 & 8.69 & 0.84 & 5.35 \\
1400 & 103924.37 & 103934.78 & 10.62 & 73.31 \\

\hline
\multicolumn{5}{c}{\textbf{maxdim: [50,100,200,400,800]}} \\
500 & 1275.64 & 1303.18 & 3.24 & 31.57 \\

\hline
\multicolumn{5}{c}{\textbf{maxdim: [100,200,400,800,1200]}} \\
1000 & 20814.49 & 23430.98 & 7.56 & 54.03 \\
1400 & 103350.87 & 103089.36 & 10.64 & 86.70 \\

\hline
\end{tabular}
\caption{Comparison of CPU and GPU runtimes for representative DMRG simulations. 
Wall time includes all runtime components such as Hamiltonian construction, 
initialization, and tensor-network optimization, while the DMRG runtime measures 
only the sweeping optimization stage.}
\label{profiling_table}
\end{table}

\begin{figure}[htbp]
\centering
\includegraphics[width=0.6\linewidth]{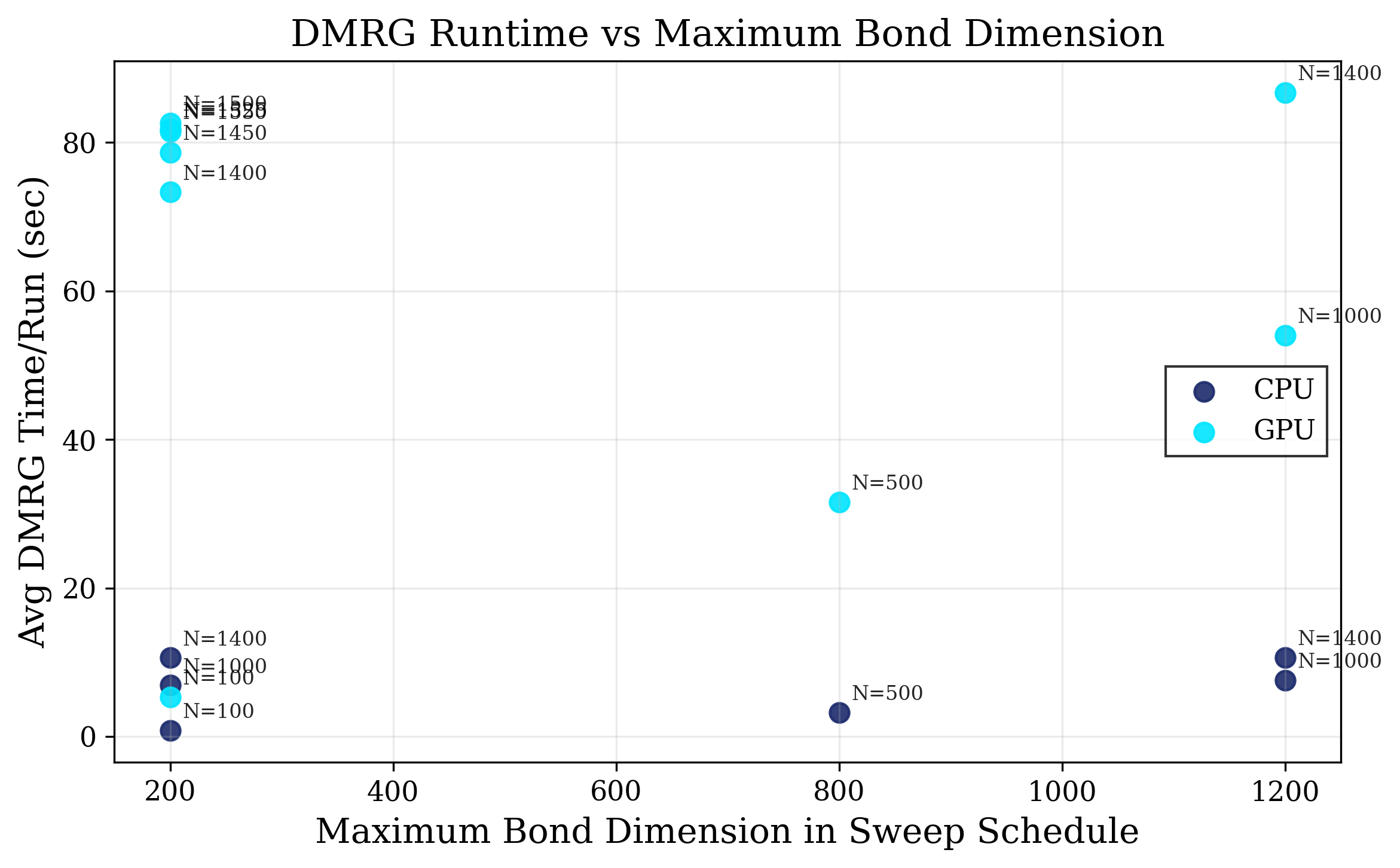}
\caption{Runtime of the DMRG optimization stage as a function of the maximum bond dimension used in the tensor-network representation for several system sizes. Each point on the plot is labeled by its system size $N$.}
\label{fig:dmrg_maxdim_scaling}
\end{figure}

We also examine how the runtime depends on a key hyperparameter of the DMRG algorithm: the maximum bond dimension used in the tensor-network representation.
Figure~\ref{fig:dmrg_maxdim_scaling} shows the DMRG-only runtime as a function of the maximum bond dimension for various $N$. As the bond dimension increases, the runtime grows substantially, reflecting the cubic scaling of tensor contractions with respect to bond dimension in matrix product state 
algorithms.

Finally, we determine the maximum system size that can be simulated within the 48-hour job limit of the Perlmutter supercomputer. Using this constraint, we find that ground-state energies can be computed for systems up to approximately $N = 1550$ particles.

\section{VQE Implementation}\label{sec:appendix-vqe}
To facilitate reproduction of our VQE baseline, we include details on the implemented optimizer and ansatz.

\subsection{Optimizer}\label{sec:appendix-vqe-optimizer}
We use the Simultaneous Perturbation Stochastic Approximation (SPSA) optimizer~\cite{spall_spsa_1992}, which estimates gradients using only two circuit evaluations per iteration regardless of parameter count, with each expectation value evaluated using 8192 shots. 

\subsection{Ansatz Selection}\label{sec:appendix-vqe-ansatz}
While \cite{Hlatshwayo_simulating_excited_states} outlines a method for choosing ansatzes in its encoding scheme, it does not trivially scale to larger dimensions ($n=N/2$). Instead, we searched for patterns in all possible (reasonable) ansatz structures as n grows. Assuming there are $n$ qubits, the maximally parametrized quantum circuit for our ansatz search would look like below where $\theta_i$ are tunable parameters.

\[
\begin{array}{c}
\Qcircuit @C=1em @R=.7em {
    & \gate{RY(\theta_1)} & \ctrl{1} & \qw & \gate{RY(\theta_{2n})} & \qw \\
    & \gate{RY(\theta_2)} & \gate{RY(\theta_{n+1})} & \ctrl{1} & \gate{RY(\theta_{2n+1})} & \qw \\
    & & & & & \\
    & \cdots & \cdots & \cdots & \cdots \\
    & \gate{RY(\theta_{n})} & \qw &  \gate{RY(\theta_{n+1 + n-1 -1})} & \gate{RY(\theta_{2n+n-1})} & \qw
}
\end{array}
\]

First, we decided that a reasonable ansatz is one consisting of 3 layers. One layer applies $R_y$ gates. The second applies $CR_y$ on adjacent qubits, and the third is structured like the first. The number of possible ansatzes fulfilling this structure is $2^{n} * 2^{n-1} * 2^n$. For $n \in [4,20]$, we examined all possible ansatzes and VQE's accuracy with the uncompressed Hamiltonian's ground eigenvalue. 

As an example, here is the table for $n=6$. The energy values were obtained for the compressed Hamiltonian with $m$-parity equal to 0 (even number of excited states):

\begin{table}[htbp]
    \centering
    \begin{tabular}{c|c|c}
        Ansatz & Energy & Relative Error \\
        \hline
         $RY_0$ & -5.189 & -0.2588 \\
         $RY_0, RY_1$ & -6.502 & -0.0712 \\
         $RY_0 ; RY_{0,1} $ & -5.101 & -0.2713 \\
         $RY_0, RY_1 ; RY_{0,1} $ & -6.502 & -0.0712 \\
         $RY_0 ; RY_0 $ & -5.101 & -0.2713 \\
         $RY_0 ; RY_{0,1};  RY_0 $ & -5.101 & -0.2713 \\
         $RY_0, RY_1; RY_0 $ & -6.499 & -0.0715 \\
         $RY_0, RY_1 ; RY_{0,1} ; RY_0 $ & -6.492 & -0.0725 \\
         $RY_0 ; RY_{0,1};  RY_1 $ & -6.503 & -0.071 \\
         $RY_0 ; RY_{0,1}; RY_0, RY_1 $ & -6.502 & -0.0712 \\
         $RY_0, RY_1; RY_1 $ & -6.5 & -0.0714 \\
         $RY_0, RY_1; RY_0, RY_1 $ & -6.505 & -0.0707 \\  
         $RY_0, RY_1; RY_{0,1}; RY_1 $ & -6.497 & -0.0719 \\
         $RY_0, RY_1; RY_{0,1}; RY_0, RY_1 $ & -6.505 & -0.0707 
    \end{tabular}
    \caption{List of all ansatzes complying with the 3-layered structure for the brute-force ansatz search. Semi-colons separate layers, and commas separate gates within a layer. All gates are either rotations about the y-axis (RY) or controlled rotations (CRY). The subscripts indicate the qubits involved with the first and second numbers corresponding to the control and target respectively. The Error column compares the compressed Hamiltonian's true ground state value with the estimated ($\frac{\epsilon_{real} - \epsilon_{compressed}}{\epsilon_{compressed}}$).}
    \label{tab:ansatz_analysis_6}
\end{table}

After experimentation on noiseless Qiskit simulators, we determined (unsurprisingly) that the optimal ansatz tended to have more gates in each layer (i.e. more parameters; see bottom rows of Table \ref{tab:ansatz_analysis_6}). With more parameters or more expressivity, a circuit has higher chance of aligning with the ground state.

\begin{figure}
    \centering
\includegraphics[width=.4\linewidth]{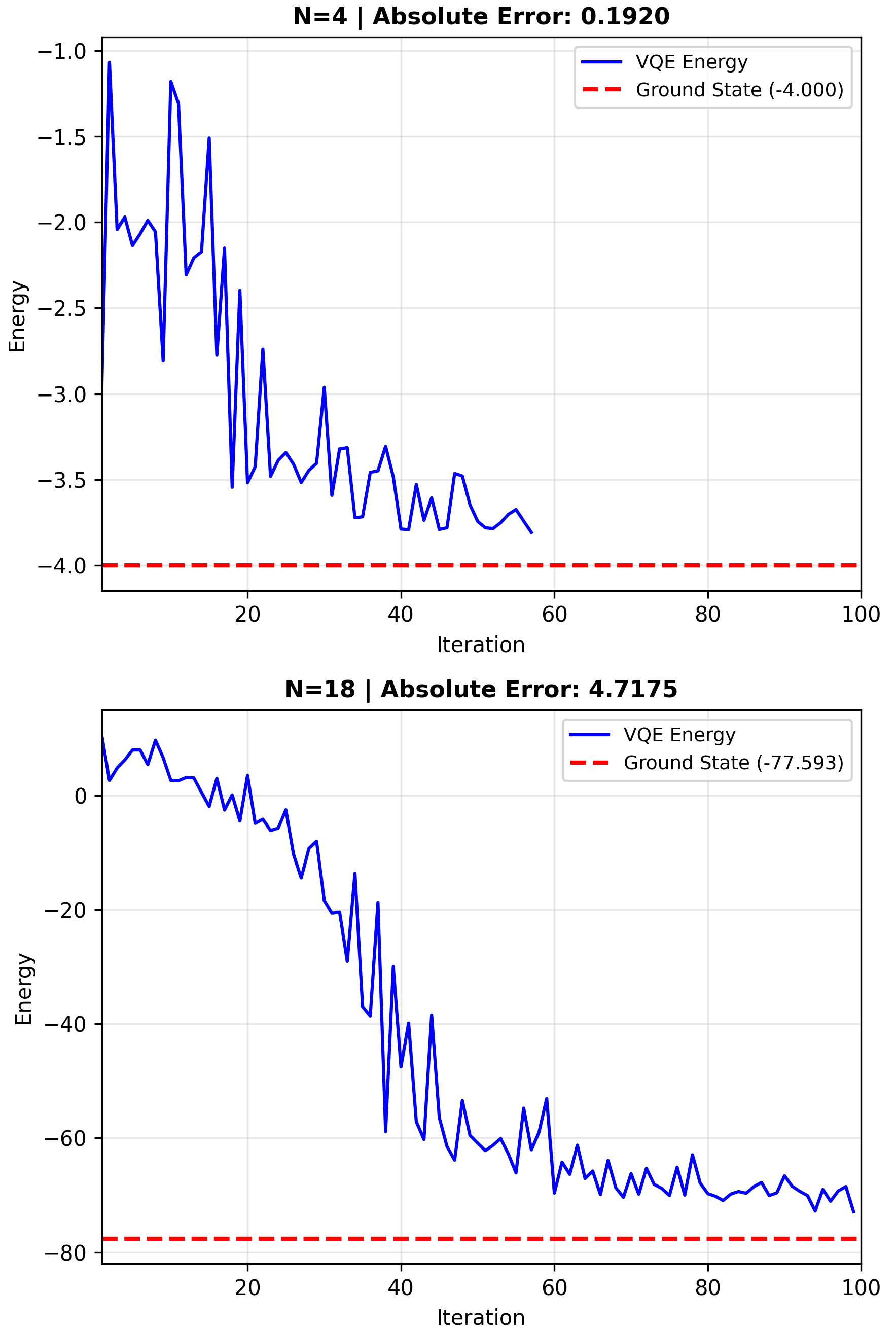}
\includegraphics[width=.4\linewidth]{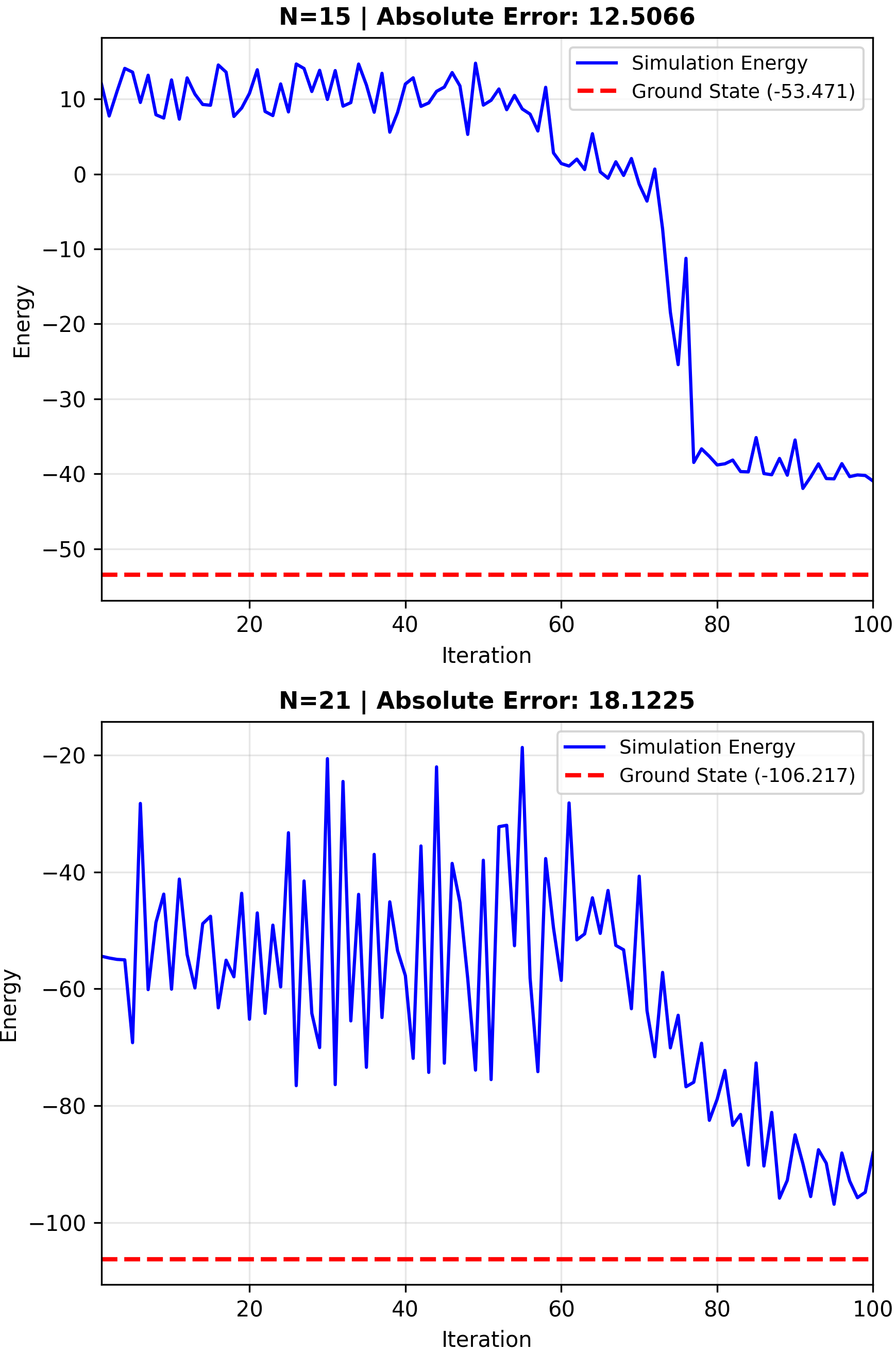}
    \caption{Iterations to convergence across four problem sizes: $N = 4,15,18,21$. All optimizer settings are the same (i.e. SPSA).}
    \label{fig:vqe_convergence}
\end{figure}

Figure~\ref{fig:vqe_convergence} examines the optimizer convergence behavior across two even and two odd problem sizes: $N = 4, 15, 18, 21$. At $N=4$, the energy descends smoothly and approaches the ground state within a modest number of iterations. At larger $N$, convergence becomes increasingly erratic. The optimizer takes significantly longer to begin descending and exhibits large fluctuations throughout, particularly for odd $N$. The $N=21$ curve never fully stabilizes within 100 iterations, suggesting that the optimizer struggles to navigate the higher-dimensional parameter landscape of the compressed Hamiltonian at these sizes. This behavior is consistent with the elevated errors seen in Figure~\ref{fig:vqe_error} for odd $N$, and suggests that alternative optimizers or problem-specific initialization strategies may be needed to improve performance in this regime.

\section{SQD Implementation}\label{sec:appendix-sqd}
We provide additional details on SQD, including the preparation of Dicke states and the scaling behavior of circuit resources.

\subsection{Dicke States}\label{sec:sqd-dicke-states}
In the case of the LMG model, the total number of computational basis states is exactly $2^{N-1}$. These states are further broken down into $\lceil \frac{N}{2} \rceil$ Dicke states. The number of computational basis states within each Dicke state grows via the binomial coefficient

\begin{equation}
     \binom{n}{k} = \frac{n!}{k!(n-k)!} 
\end{equation}

\begin{figure}[!ht]
    \centering
    \hspace*{0\linewidth}
    \includegraphics[width=0.5\linewidth]{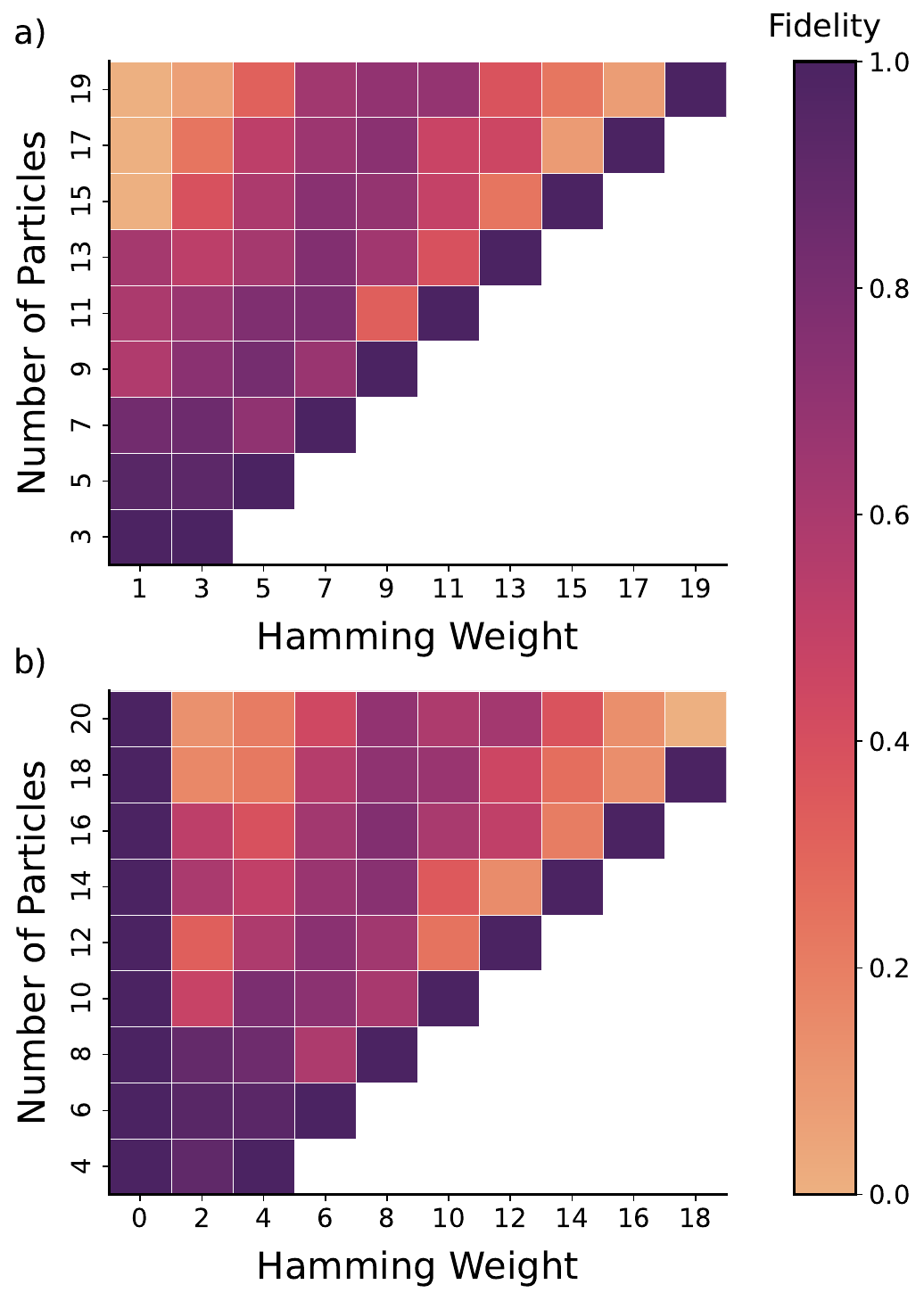}
    \caption{fidelity between the target Dicke state for given $N$ and Hamming weight  $k$, and the experimental measurement distribution.}
    \label{fig:heatmap}
\end{figure}

This scaling motivated the use of a weighted distribution to sample an ensemble of Dicke states for a given $N$ since the number of basis states grows as you approach $n = k/2$.  However, due to the well-defined symmetry of Dicke states, the self-configuration recovery algorithm is capable of correcting any corrupted bitstring into a valid one by simply flipping bits until the desired hamming weight is achieved. While good for the purposes of increasing the fidelity of the experiment, this can obscure the intrinsic performance of the quantum processor. To address this, we evaluate the state fidelity prior to recovery, allowing for a transparent assessment of the processor's ability to prepare the Dicke state circuits using this weighted distribution shown in Figure~\ref{fig:heatmap}. 

\subsection{Circuit Resource Scaling}\label{sec:sqd-circuit-scaling}

\begin{figure}[!ht]
    \centering
    \includegraphics[width=0.5\linewidth]{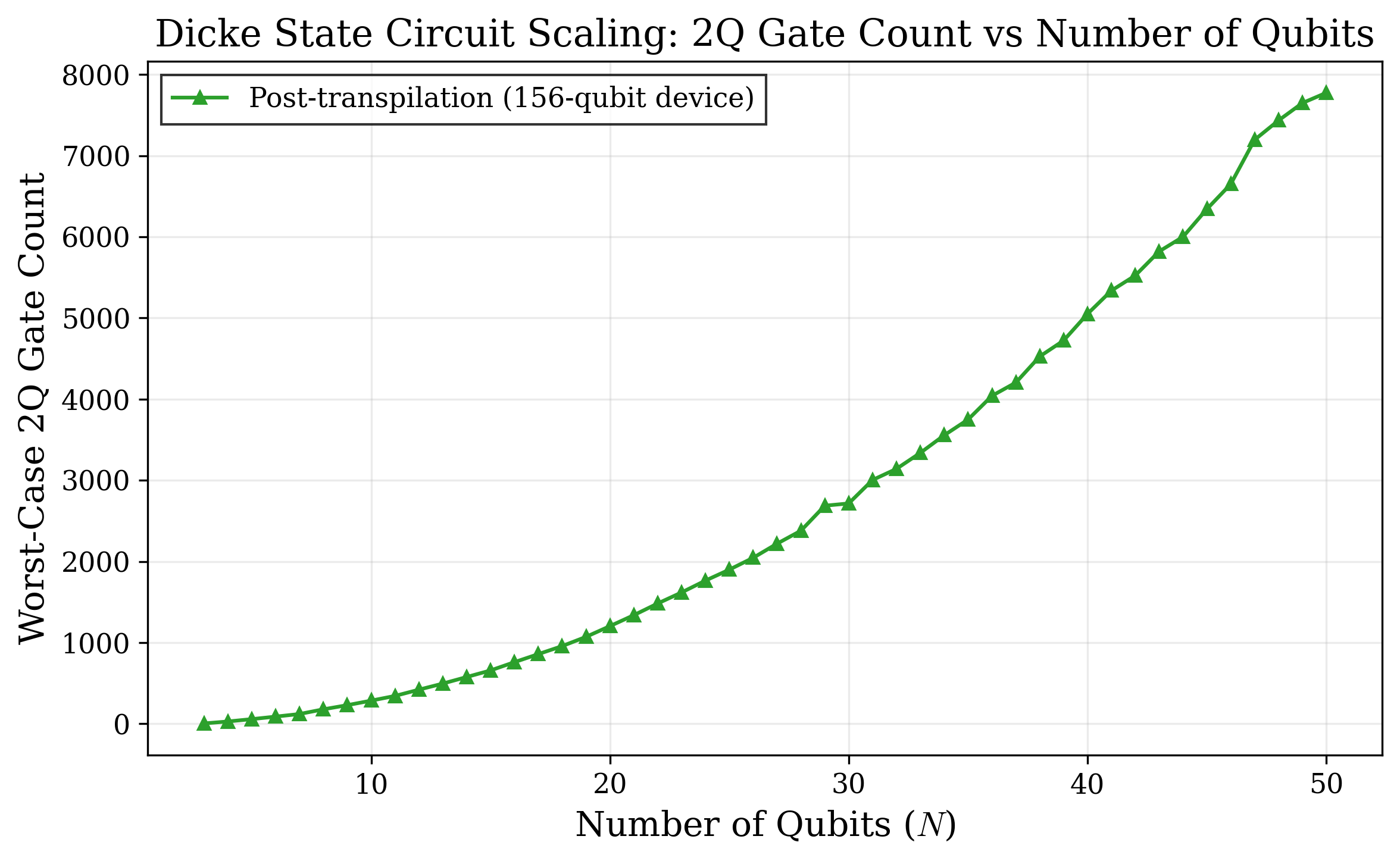}
    \caption{Worst-case 2Q gate count of $|D_N^k\rangle$ after transpilation to a IBM heavy-hex device (optimization level~3, best of 20 transpiler seeds), selecting $k = \lfloor N/2 \rfloor$ for each $N$.}
    \label{fig:depth}
\end{figure}

\begin{figure}[!ht]
    \centering
    \includegraphics[width=0.5\linewidth]{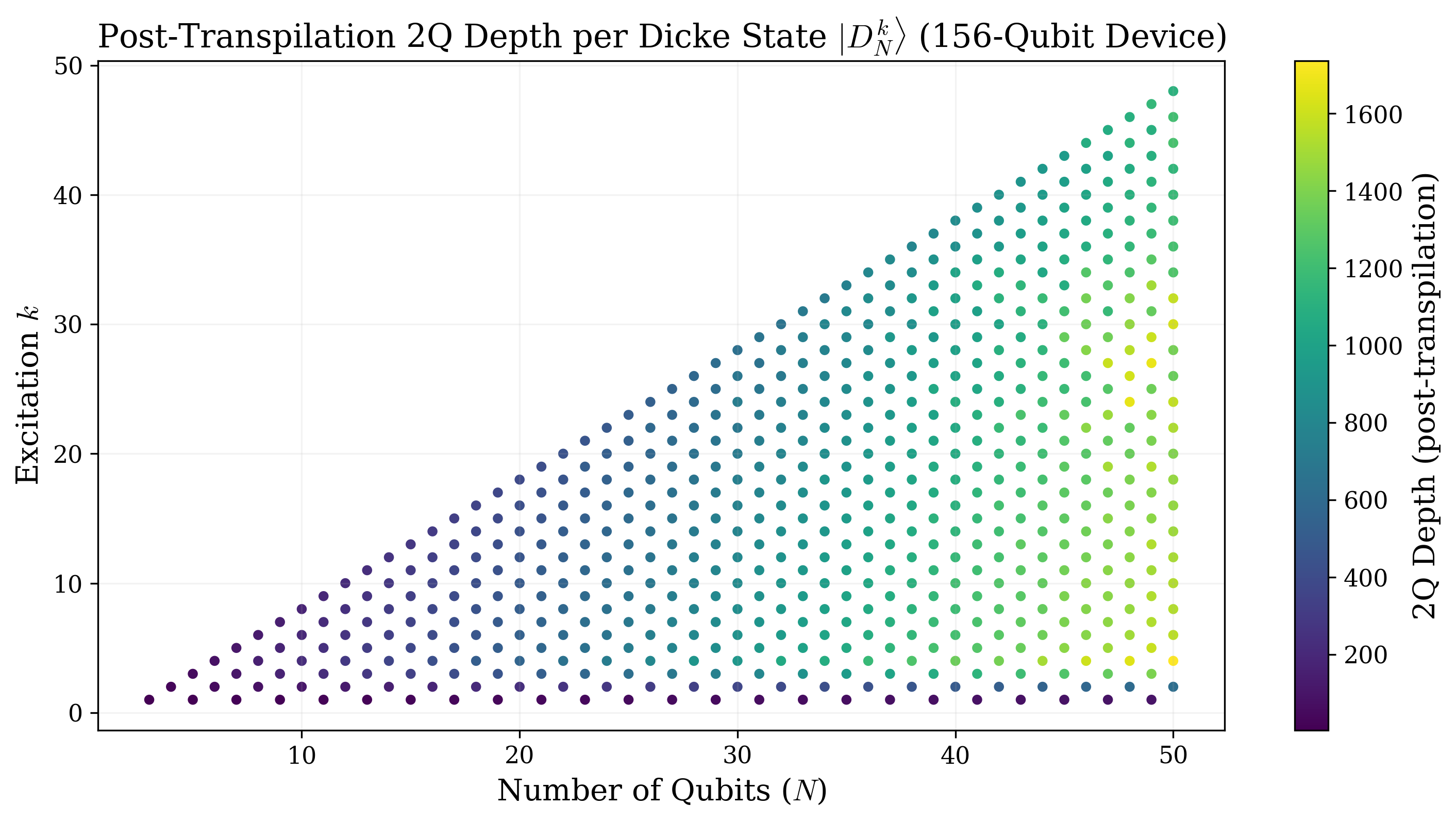}
    \caption{Two-qubit circuit depth of each $|D_N^k\rangle$ after transpilation to a IBM heavy-hex topology (optimization level~3, best of 20 transpiler seeds), with color encoding the optimized 2Q depth per $(N, k)$ pair.}
    \label{fig:depth_heatmap}
\end{figure}

Beyond accuracy, a key figure of merit for near-term deployability is circuit resource cost. The worst-case 2Q gate count, occurring at $k = \lfloor N/2 \rfloor$, follows $\mathcal{O}(k(2N-k)) \sim \mathcal{O}(N^2)$, reaching approximately 8000 gates at $N = 50$ after transpilation to the native gate set of the IBM heavy-hex device, as shown in Figure~\ref{fig:depth}. The circuit depth heatmap in Figure~\ref{fig:depth_heatmap} reveals that this resource cost is strongly concentrated near $k \approx N/2$: circuits preparing low- or high-excitation Dicke states ($k \ll N/2$ or $k \gg N/2$) are substantially shallower. This structure suggests a natural optimization strategy: weighting the shot budget toward moderate-excitation Dicke states while deprioritizing the deepest circuits simultaneously improves subspace coverage and reduces average circuit depth. The maximum observed depth of ${\sim}1700$ at $N = 50$ already approaches the coherence limits of current devices, reinforcing that the shot-budget constraint and circuit depth jointly define the practical scaling ceiling for SQD on near-term hardware.

Taken together, these results establish that SQD achieves near-exact accuracy for the LMG model in the regime $N < 20$, where the symmetry sector is fully coverable by available shot budgets and circuit depths remain manageable. Extending SQD to larger system sizes will require either substantially larger shot budgets, improved subspace sampling strategies, or shallower circuit constructions that reduce the impact of hardware noise.

\end{document}